\documentclass[10pt,journal]{IEEEtranTCOM}

\usepackage[english]{babel}
\usepackage[usenames]{color}
\usepackage[cp1250]{inputenc}
\usepackage{amsfonts}
\usepackage{amsthm}
\usepackage{graphicx}
\usepackage{epsfig}
\usepackage{mathrsfs}
\usepackage{amsmath}
\usepackage{algorithm}
\usepackage{algorithmic}
\usepackage{hyperref}

\theoremstyle{plain}

\begin{document}
\newcommand{\bea}{\begin{eqnarray}}
\newcommand{\eea}{\end{eqnarray}}
\newcommand{\be}{\begin{equation}}
\newcommand{\ee}{\end{equation}}
\newcommand{\beas}{\begin{eqnarray*}}
\newcommand{\eeas}{\end{eqnarray*}}
\newcommand{\bs}{\backslash}
\newcommand{\bc}{\begin{center}}
\newcommand{\ec}{\end{center}}
\def\SC {\mathscr{C}}

\title{Four-dimensional understanding\\ of quantum mechanics and Bell violation}
\author{\IEEEauthorblockN{Jarek Duda}\\
\IEEEauthorblockA{Jagiellonian University,
Golebia 24, 31-007 Krakow, Poland,
Email: \emph{dudajar@gmail.com}}}
\maketitle

\begin{abstract}
While our natural intuition suggests us that we live in 3D space evolving in time, modern physics presents fundamentally different picture: 4D spacetime, Einstein's block universe, in which we travel in thermodynamically emphasized direction: arrow of time. Arguments for such nonintuitive and nonlocal living in kind of "4D jello" come among others from: Lagrangian mechanics we use from QFT to GR saying that history between fixed past and future situation is the one optimizing action, special relativity saying that different velocity observers have different own time directions, general relativity deforming shape of the entire spacetime up to switching time and space below the black hole event horizon, or the CPT theorem concluding fundamental symmetry between past and future for example in the Feynman-Stueckelberg interpretation of antiparticles as propagating back in time.

Accepting this nonintuitive living in 4D spacetime: with present moment being in equilibrium between past and future - minimizing tension as action of Lagrangian, leads to crucial surprising differences from intuitive "evolving 3D" picture - allowing to conclude Bell inequalities, violated by the real physics. Specifically, particle in spacetime becomes own trajectory: 1D submanifold of 4D, making that statistical physics should consider ensembles like Boltzmann distribution among entire paths (like in Ising model), what leads to quantum behavior as we know from Feynman's Euclidean path integrals or similar Maximal Entropy Random Walk (MERW). It results for example in Anderson localization, or the Born rule with squares - allowing for violation of Bell inequalities. As e.g. for S-matrix, quantum amplitude turns out to describe probability at the end of half-spacetime from a given moment: toward past or future, to randomly get some value of measurement we need to "draw it" from both time directions - getting the squares of Born rules. Tension from both time directions is also suggested in quantum experiments like Wheeler's delayed choice experiment, it will be argued that it is also crucial in quantum algorithms like Shor's, there are also suggested hypothetical better alternatives.
\end{abstract}
\textbf{Keywords:} quantum mechanics, nature of time, spacetime, Einstein's block universe, Born rule, Bell inequalities, Shor algorithm, Euclidean path integrals, statistical physics, Ising model, \href{https://en.wikipedia.org/wiki/Maximal_entropy_random_walk}{Maximal Entropy Random Walk} (MERW)
\section{Introduction}
\begin{figure}[t!]
    \centering
        \includegraphics{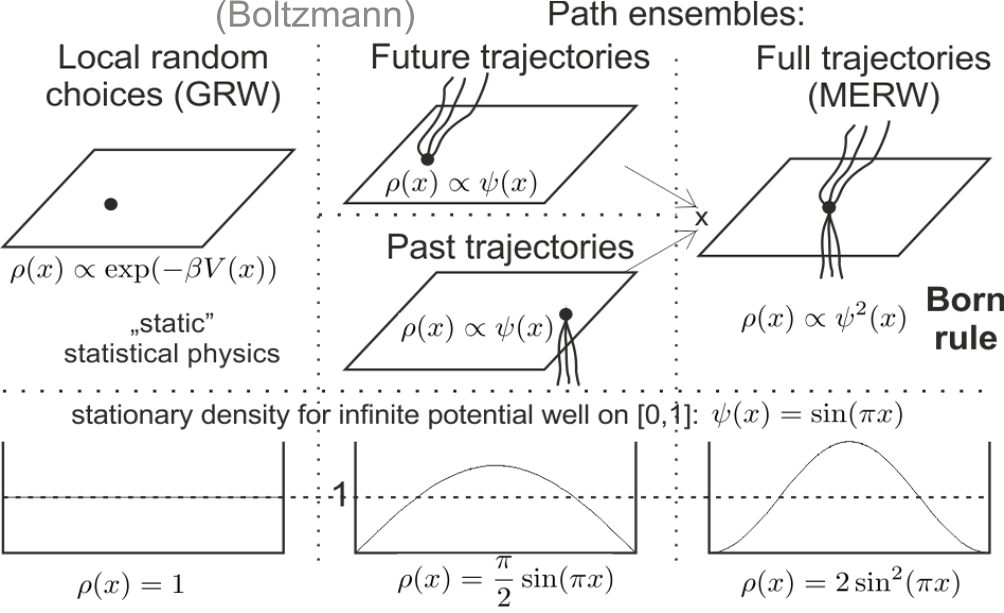}
        \caption{Three philosophies for finding probability distribution of a walker, for example electron. Left: "static 3D" situation. Center: example of "evolving 3D" philosophy of time, assuming Boltzmann distribution among trajectories from the past to the current moment. As it will be discussed, it leads to probability distribution proportional to the first power of quantum ground state amplitude. Right: "4D spacetime" situation where particle becomes its full trajectory. Like in Feynman's Euclidean path integrals (or Ising model), Boltzmann distribution among such full trajectories leads to Born rule: focusing on a fixed-time cut, to randomly get a given position, we need "to draw it" from both past and future half-trajectories. Hence, probability is proportional to the product of two (identical here) "evolving 3D" probabilities, getting the Born rule. Bottom: stationary probability distribution predicted by them for $[0,1]$ infinite potential well, only the "spacetime" consideration agrees with the QM ground state. \newline}
       \label{born}

        \includegraphics{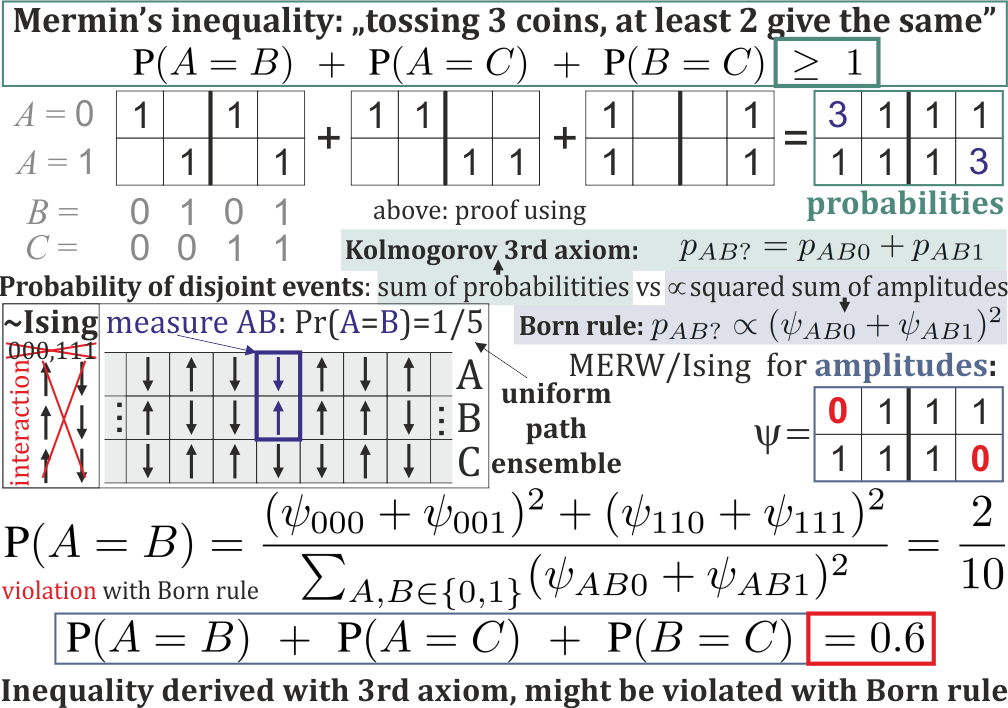}
        \caption{Top: schematic proof of simple Bell-like inequality~(\cite{Bell,mermin}): assuming any hidden probability distribution among $2^3=8$ possibilities $\{0,1\}^3$ for 3 binary variables ABC, we get inequality written at the top. Surprisingly, QM allows to violate it. Bottom: example of its violation using Born rules like in MERW e.g. in Ising model (Boltzmann vs Feynmnan path ensemble): as in Fig. \ref{born} amplitude corresponds to probability toward one of two directions (spatial in Ising, temporal in QM), to randomly get some value we need to "draw it" twice, getting probability as normalized square of sum of amplitudes. To realize it with Ising model, as shown we would need width 3 lattice of spins with interaction preventing $\uparrow\uparrow\uparrow$ and $\downarrow\downarrow\downarrow$ triples, and measure 2 out of 3 spins in a triple. Details in Section \ref{bellsect}.}
        \label{bell}
\end{figure}

\begin{figure}[t!]
    \centering
        \includegraphics{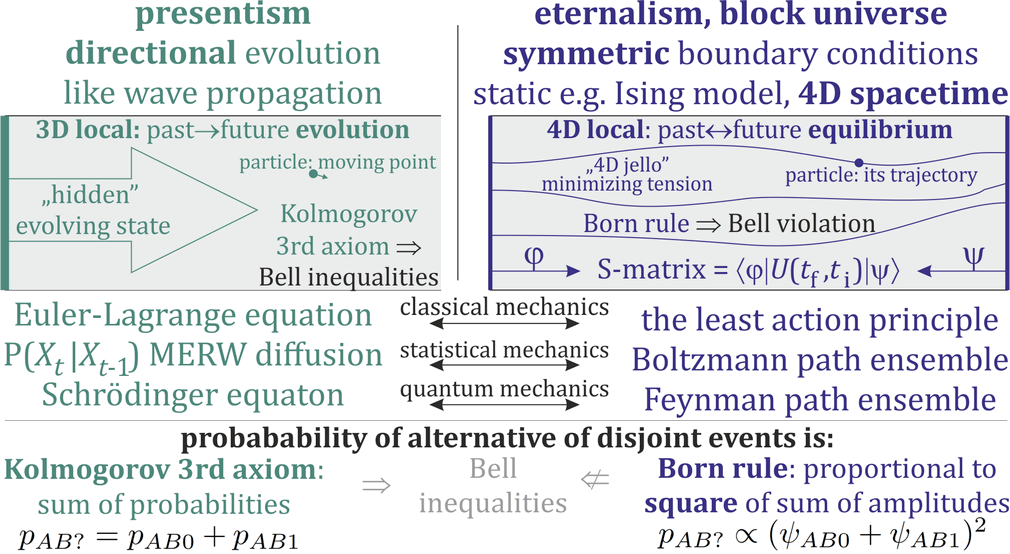}
        \caption{Two philosophies of time: \href{https://en.wikipedia.org/wiki/Philosophical_presentism}{presentism} evolving the current 3D situation, and \href{https://en.wikipedia.org/wiki/Eternalism_(philosophy_of_time)}{eternalism}, block universe imagining travel through some already chosen 4D solution as past-future equilibrium - like in "4D jello", shape of 4D spacetime in the general relativity (trying "to develop" spacetime with Euler-Lagrange equations seems highly problematic). We can translate between such two types of solutions e.g. in classical, quantum, statistical mechanics. However, solutions originally found with one of them have subtle differences, 3D locality suggested by our intuition is different from 4D locality. For example while directional solutions should satisfy Bell inequalities, symmetric ones allow to replace Kolmogorov 3rd axiom with Born rule, allowing for violation of Bell-like inequalities. We will focus here on MERW as diffusion chosen accordingly to the maximal entropy principle - as required by statistical mechanics models, also used in Ising model. }
       \label{overv}
\end{figure}

\begin{figure}[b!]
    \centering
        \includegraphics{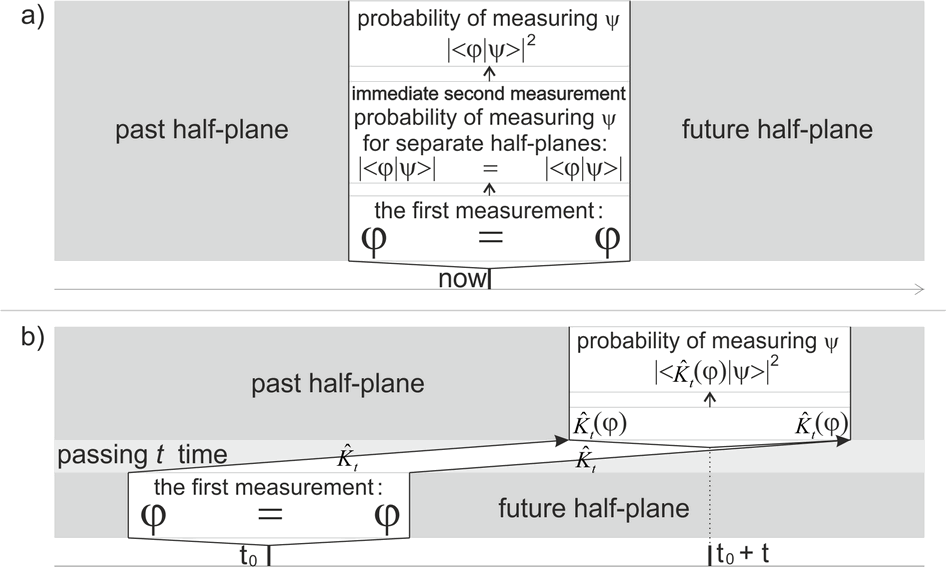}
        \caption{The lesson from Fig. \ref{born} taken to quantum evolution and measurement: in the "spacetime" philosophy, quantum amplitude describes not only probability distribution at the end of the past, but simultaneously also at the beginning of the future. To calculate probability of some value of measurement in a fixed time cut, we need to multiply the (identical) predicted probabilities for the past and future directions, getting Born rule: squares known in QM formalism, which lead e.g. to Bell violation. }
       \label{born1}
\end{figure}

Starting with special relativity (SR) a century ago, modern physics uses 4D spacetime view of our world - Einstein's block universe, in which we travel in time direction. Also a century ago quantum mechanics (QM) was born, bringing many nonintuitive consequences, like violation of Bell inequalities. As briefly presented in Fig. \ref{born}, \ref{bell}, \ref{born1} and \ref{overv}, this article argues that these two revolutions of our understanding - violating our natural intuitions, are in fact deeply connected: that living in spacetime has surprising microscopic consequences seen in QM formalism. Let us start with reminding well known arguments and consequences of the spacetime view of modern physics.

Our natural human intuition has evolved for past-future reason-result chains of consequences: initiated in our Big Bang, leading to us through creation of our planet, evolution, our development. However, since the special relativity we know that "situation in a given moment", more formally called the hypersurface of the present, in fact depends on the observer's frame of reference: it changes with his velocity accordingly to Lorentz boost, which also modifies the direction of time. The general relativity takes it even further, modifying the entire spacetime accordingly to local mass/energy concentrations, up to extreme situation below the black hole event horizon, where time and space directions literally switch places, making "situation in a given moment" very far from our biological intuition.

This ambiguity of time direction is also seen in the Lagrangian formalism we successfully use to describe reality in all scales: from quantum field theories to the general relativity. It has multiple equivalent formulations, starting with our intuitive "evolving 3D" picture: Euler-Lagrange equation allows to evolve the situation forward in time, from a situation (as values and the first derivatives) in a given moment. However, mathematically it also allows to use these equations to evolve situation backward in time, as Lagrangian mechanics is usually time (or CPT) symmetric. Quite different  formulation is through action optimization: fixing situation (only values, without the first derivatives) in two moments in time, the history between them is the one optimizing action. While we can translate between such solutions, those originally found with each of them have subtle differences, visualized in Fig. \ref{overv}, e.g. action optimization has a different  locality than the one assumed in Bell theorem.

\begin{figure}[t!]
    \centering
        \includegraphics{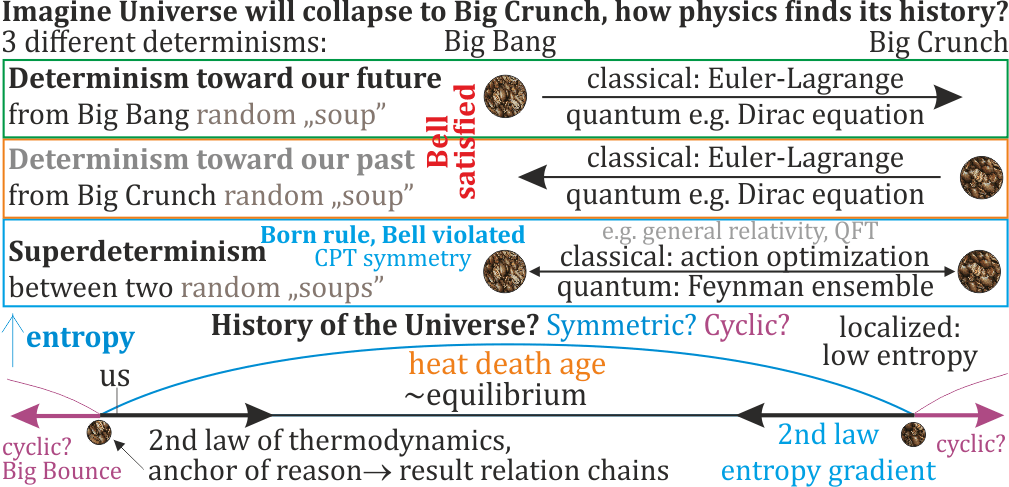}
        \caption{Top: three philosophies of determinism, for simplicity assuming that our Universe will finally collapse into Big Crunch - when again everything will be localized, hence entropy should be low. Intuitive presentism philosophy imagines evolution from some random soup in our Big Bang. Mathematically we could switch sign of time $t\to -t$ in equations, and consider evolution e.g. from having similar parameters random soup of Big Crunch. We could also use symmetric etetrnalism/block universe philosophy with single least action history of the Universe, or their Feynman ensemble. In such 4D optimization or ensemble, influence of e.g. measurement is in both time directions, making such solution \href{https://en.wikipedia.org/wiki/Superdeterminism}{superdeterministic} - in contrast to the first two, not necessarily satisfying Bell inequalities. With two amplitudes in Born rule coming from propagators from two time directions as e.g. in S-matrix. Bottom: a hypothetical symmetric sketch of history of the Universe. Similarity between Big Bang and Big Crunch suggests they should have similar low entropy due to localization, enforcing local entropy gradients - 2nd law of thermodynamics pointing out nearby collapsed state, which are also anchors starting reason-result relation chains leading to stars, planet formation, evolution of life. Such opposite evolutions would be separated by thermal death age, destroying (nearly?) all low entropic objects from hypothetical opposite evolution. CPT symmetry suggests symmetric counterparts: Big Bounces, in the diagram starting another two histories, and further leading to \href{https://en.wikipedia.org/wiki/Cyclic_model}{cyclic Universe} hypothesis.}
       \label{cyclic}
\end{figure}

Lagrangian mechanics for field theories can additionally be Lorentz invariant: compatible with Lorentz boost change of direction of time. To get some intuition, let us briefly remind the simplest scalar field theory: with Hamiltonian (energy density) and Lagrangian:
 $$\mathcal{H}=\frac{1}{2}\sum_{\nu=0,1,2,3} (\partial_\nu \phi)^2 + V(\phi)$$
 $$  \mathcal{L}=\frac{1}{2}\left( (\partial_0 \phi)^2-\sum_{i=1,2,3} (\partial_i \phi)^2 \right) - V(\phi) $$
Surprisingly, energy density (Hamiltonian) is often completely 4D symmetric like here: does not emphasize any time direction in 4D. Choosing a frame of reference, it determines time direction '0', for which we can find the Lagrangian which Legendre transform is the given Hamiltonian $(\pi = \partial_0 \phi,\ \mathcal{L}=\pi\, \partial_0 \phi - \mathcal{H})$. This Lagrangian emphasizes the chosen time direction. To summarize, energy density (Hamiltonian) of a Lorentz invariant field theory often allows to imagine the spacetime as completely symmetric "4D jello": minimizing tension as Hamiltonian. Choosing some time direction and situation in its hypersurface of the present, we can transform Hamiltonian to Lagrangan, find Euler-Lagrange equation for it, and use it to evolve situation from this arbitrary hyperplane in a chosen direction of time.

Fundamental similarity of past and future is also the base of quantum field theories requiring \href{https://en.wikipedia.org/wiki/CPT_symmetry}{CPT symmetry} due to the Schwinger's CPT theorem~\cite{CPT}. Feynman diagrams represent antiparticles as propagating backward in time (Feynman-Stueckelberg interpretation)~\cite{feynman}.\\

\begin{figure}[t!]
    \centering
        \includegraphics{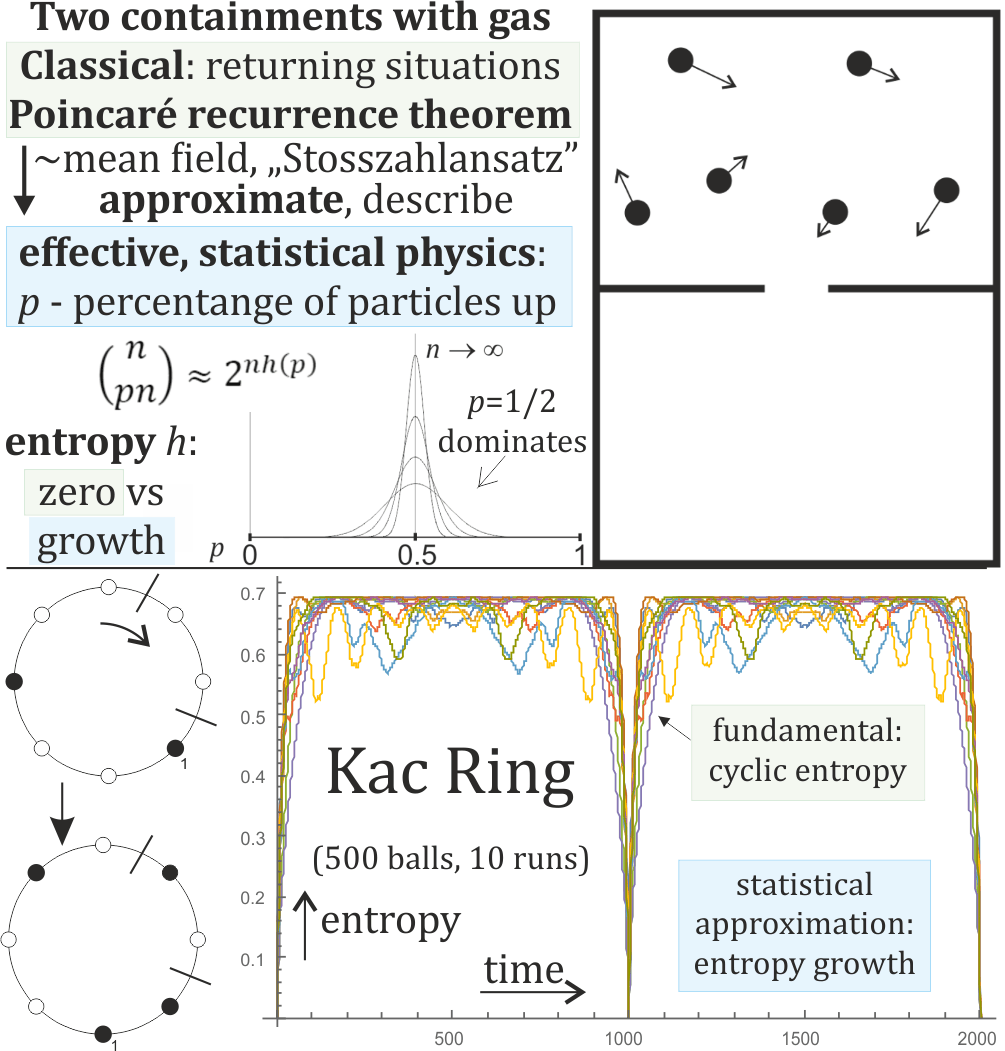}
        \caption{Time symmetry vs entropy growth in some time-symmetric systems. Entropy e.g. Shannon: $h(p)=-p\lg(p)-(1-p)\lg(1-p)$ is nonzero only for nontrivial probability distributions, for deterministic situations it is zero. Top: two containments with gas examples - using classical particle, situation is returning due to \href{https://en.wikipedia.org/wiki/Poincar\%C3\%A9_recurrence_theorem}{Poincare recurrence theorem}, entropy is zero. However, using "$p$-the number of particles up" as the only description, what is mean-field approximation also called Stosszahlansatz, we get entropy growth as the number of possibilities/combinations $\binom{n}{pn}\approx 2^{nh(p)}$ is maximized for $p=1/2$. 
        Bottom: \textbf{Kac ring}~\cite{kac}: imagine a ring with $n$ white and black balls, and there are chosen some marked positions (two drawn segments). In each step all balls rotate by one position, each ball going through a marked position flips color white$\leftrightarrow$black. Denoting $p\in[0,1]$ as the current number of white balls divided by $n$. A natural looking assumption (Stosszahlansatz) is that this $p$ also describes proportion of colors of balls before the marked positions - this assumption allows to conclude that $p\to 1/2$ to maximize the entropy. However, if starting from all white balls and performing $n$ steps, all balls go through the same number of flips, returning to the zero entropy situation. The plots show entropy for 10 simulations with $n=500$ balls and randomly chosen marked positions with 0.01 probability. Analogously, for also time-symmetric classical mechanics, the Stosszahlansatz used in Boltzmann \href{https://en.wikipedia.org/wiki/H-theorem}{H-theorem}~\cite{bolt} to conclude entropy growth is assumption (mean field-like approximation) that two particles participating in a collision have independently randomly chosen energies, directions and starting points.}
       \label{kac}
\end{figure}

In contrast, against e.g. SR and fundamental CPT symmetry, 2nd law of thermodynamics emphasizes some "arrow of time". However, thermodynamics is not fundamental, only effective modelling: describes the most probable statistical behavior - accordingly to (Jaynes') \href{https://en.wikipedia.org/wiki/Principle_of_maximum_entropy}{principle of maximum entropy}, averaging over unknowns. While physics fundamentally suggests quite symmetric "4D jello", this symmetry is clearly broken on thermodynamical level in the actual solution we live in. Like a fundamentally symmetric water surface can obtain a state (solution) with this symmetry broken e.g. by throwing a rock. As proven for example in \href{https://en.wikipedia.org/wiki/H-theorem}{Boltzmann H-theorem}~\cite{bolt}, entropy growth is a natural statistical tendency even for time-symmetric models, what seems self-contradictory as after applying such symmetry, the same proof should conclude opposite entropy growth. Figure \ref{kac} presents a simple Kac model which gives a valuable lesson about this apparent paradox - such proofs of entropy growth have to rely on looking natural assumptions of some uniformity (called \textbf{Stosszahlansatz}), allowing to break time-symmetry of the model. However, hidden structure of such time-symmetric system can also lead to entropy decrease and bouncing from zero entropy as in the plot in Fig. \ref{kac}. The direct reason for entropy growth with our time arrow might be our Big Bang: having all matter localized and so low entropy, starting the cascade of reason-result chains of consequences leading to our current situation. Assuming our Universe will finally gravitationally collapse and bounce starting new Universe, as in Fig. \ref{cyclic}, the fundamental CPT symmetry of our physics suggests that such Big Bounce situation might be also nearly symmetric from the point of view of entropy, like in bounces for Kac ring in Fig. \ref{kac}.\\

\begin{figure}[t!]
    \centering
        \includegraphics{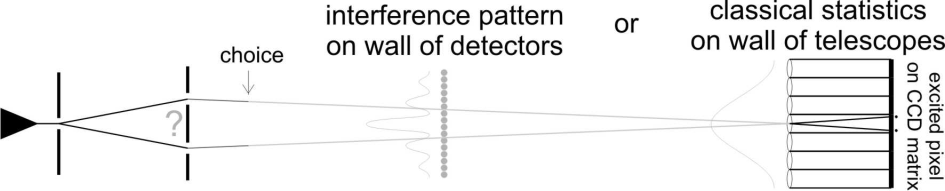}
        \caption{\href{https://en.wikipedia.org/wiki/Wheeler\%27s_delayed-choice_experiment}{Wheeler's delayed-choice experiment} as one of QM examples suggesting future-past causality: imagine there is some double-slit type of experiment in a distance, for example light from a star can pass near one of sides of a planet on the way. Later (by travel time) we choose how to observe it: if resolution of our telescope allows to distinguish the two slits we get corpuscular behavior, otherwise we get wave behavior: quantum interference. Hence it seems like in the future we can choose between wave or particle past nature of photons. In experimental confirmation~\cite{wheeler} there was used Mach-Zehnder setting: choosing between classical and quantum photon behavior (lifting or not the last half-silvered mirror) after it went through the first half-silvered mirror. There are also more sophisticated QM experiments suggesting future-past causality, like the delayed choice quantum erasure~\cite{walborn} or experiment from "Asking photons where they have been" article~\cite{asking}.}
       \label{wheeler}
\end{figure}

While it is difficult for us to really accept, we see that especially special relativity and Lagrangian mechanics for fields provide a picture that, against our natural intuitions, we live in spacetime as kind of "4D jello" minimizing tension defined by energy density (Hamiltonian). Hence, the present moment is kind of equilibrium between past and future situation (like in time/CPT symmetry of fundamental theories we use), what makes physics nonlocal in "evolving 3D" sense (but local in 4D view). In contrast, our intuition considers only consequences from the past time direction. Nature provides many suggestions that the resulting nonintuitiveness is connected with the strangeness of quantum mechanics, like the Wheeler experiment briefly presented in Fig. \ref{wheeler}, or the Delayed Choice Quantum Erasure. For example there is John Cramer's \href{https://en.wikipedia.org/wiki/Transactional_interpretation}{transactional interpretation} of QM~\cite{cramer} based on this inherent time symmetry of quantum unitary evolution, suggesting propagation of information in both time directions.

Time symmetric formulation of QM is also advocated by Aharonov~\cite{tsvf} in so called \href{https://en.wikipedia.org/wiki/Two-state_vector_formalism}{two-state vector formalism}: seeing the present moment as a result of two propagators: from minus and plus infinity, what is to analogous to \href{https://en.wikipedia.org/wiki/S-matrix#Interaction_picture}{scattering matrix}: $\langle\psi_f|U(t_f,t_i)|\psi_i\rangle$, and the view presented here. Article "Asking photons where they have been"~\cite{asking} presents its very nice experimental conformation: vibrating mirrors allows to conclude from the final light beam which mirrors have been visited. However, in properly chosen setting they obtain also signal from mirrors which naively should not be visited, unless we focus exactly on mirrors visited by photons propagating in both time directions.\\

\begin{figure}[t!]
    \centering
        \includegraphics{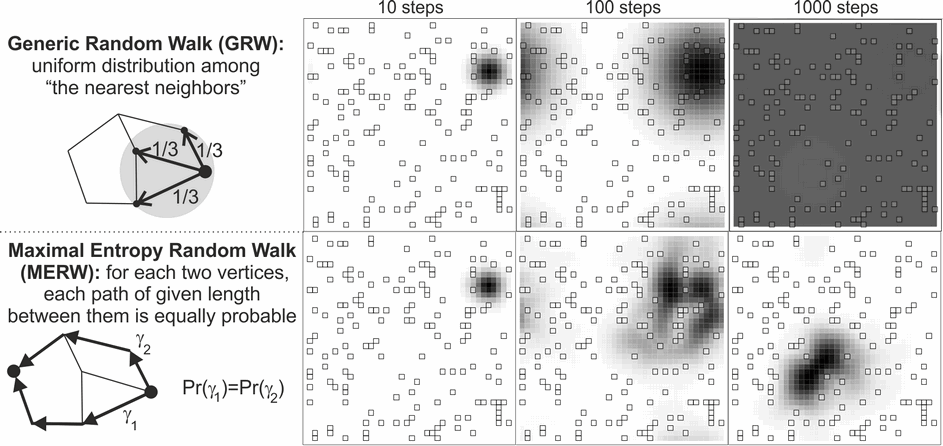}
        \caption{Left: GRW and one of formulations of MERW for a simple graph. Right: example of comparison of their evolution for a defected $40\times 40$ lattice with cyclic boundary conditions - all but the marked vertices have self-loop: edge to itself. Stationary probability density of GRW is proportional to degree of vertex, getting very weak localization: defects have $4/5$ probability of the remaining nodes. In contrast, while MERW seems to have similar local behavior (density after 10 steps), it leads to a completely different stationary density, exactly like for the quantum ground state: square of coordinates of the dominant eigenvector of minus Hamiltonian (simplified Bose-Hubbard here: $\mathcal{H}=-\sum_{\textrm{edge }(ij)} a_j^+ a_i$). Such defected lattice can be seen as a simple model of semiconductor: regular lattice with rare dopants. Nearly uniform electron density predicted by standard diffusion (GRW) means that electrons should flow if attached electric field - making it a conductor against experiment. In contrast, QM and MERW predict electron prisoned e.g. in local defect-free regions (called Lifhitz spheres). Simple GRW/MERW conductance simulator is available in~\cite{cond}. MERW transition probabilities use eigenvector and so it depends on the entire system, making this model nonlocal. However, in such thermodynamical view the walker does not directly use these nonlocal probabilities, only we use them to predict the most probable evolution of its probability distribution accordingly to our knowledge. }
       \label{merw}
\end{figure}

In this article we will focus on  statistical consequences of given moment (hypersurface of the present) being in equilibrium between past and future in spacetime as kind of "4D jello". In this picture particles are no longer just "moving points", but rather their trajectories: one-dimensional submanifolds of the spacetime. From statistical physics perspective, it brings the question of what objects should we use in the considered ensembles, e.g. while assuming Boltzmann distribution like presented in Fig. \ref{born} - only the last one: considering Boltzmann distribution among full trajectories, like in Feynman's Euclidean path integrals or related Maximal Entropy Random Walk (MERW)~(\cite{merwprl, myphd}), has thermodynamical agreement with predictions of QM: leading to stationary probability distribution of the quantum ground state, with crucial differences like avoiding the boundaries for $[0,1]$ infinite potential well. Boltzmann path ensemble is also at heart of Ising model - MERW can be also used as random walk along Ising sequence.

\begin{figure}[b!]
    \centering
        \includegraphics[width=8.5cm]{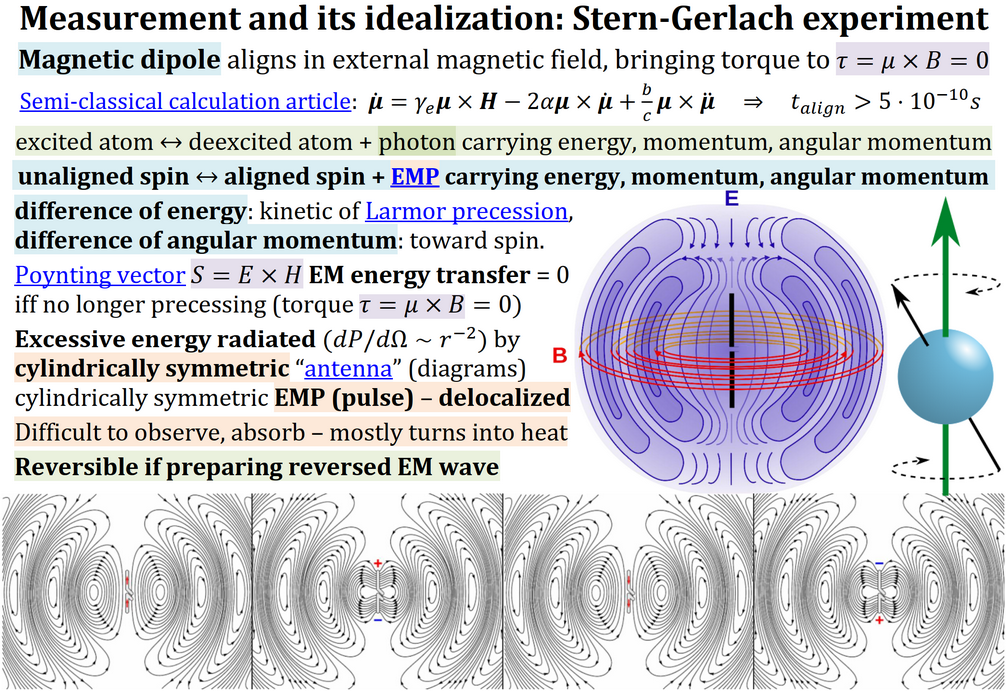}
        \caption{\href{https://en.wikipedia.org/wiki/Stern\%E2\%80\%93Gerlach_experiment}{Stern-Gerlach experiment} is often imagined as idealization of quantum measurement - starting with unknown/random directions of spins in strong magnetic field, ending with quantized: spin parallel or anti-parallel to strong external magnetic field $B$. Magnetic dipole moment $\mu$ undergoes \href{https://en.wikipedia.org/wiki/Larmor_precession}{Larmor precession} in external magnetic field due to $\tau=\mu\times B$ torque, which is zero only for the (obtained) parallel or anti-parallel alignment - no longer requiring additional kinetic energy of precession, like for excited atoms this excessive energy should be released - aligning the spin. From the other side, precessing magnetic dipole creates varying magnetic field, which radiates energy as EM waves like e.g. \href{https://en.wikipedia.org/wiki/Antenna_(radio)}{antenna}. It allows to imagine this alignment (measurement?) as example of (delocalized) \href{https://en.wikipedia.org/wiki/Electromagnetic_pulse}{electromagnetic pulse (EMP)}, which would be reversible if being able to prepare reversed EM pulse - carrying difference of energy and angular momentum between initial and final spin directions. The diagram uses images from the linked Wikipedia articles and formulas from semi-classical calculation article\cite{stern}. }
       \label{stern}
\end{figure}

In the next Section we will focus on MERW philosophy, briefly presented in Fig. \ref{merw}, as a reparation of standard diffusion models - which for example wrongly predict that semiconductor should have nearly uniform probability distribution of electrons, making it a conductor against experiment (tiny electric field would cause electron flow). This fundamental disagreement was repaired by QM showing strong localization property for these electrons (e.g. Anderson), preventing the conductance. This crucial problem of stochastic modelling has caused that it is currently seen as a completely different realm than QM. However, electrons are indivisible quants of electric charge, what should prevent them from being objectively blurred. Heisenberg uncertainty principle limits abilities of measurement, which are sophisticated destructive processes idealized for example by the \textbf{Stern-Gerlach experiment} - having also reversible interpretations like in Figure \ref{stern} . In contrast, this principle is commonly seen as limitation for example for objective position of electron in atom. However, modern techniques like field-emission electron microscopy already allow to get resolution below the size of atomic orbital: strip electrons from single atom, use EM field shaped to act as a lens for magnification, and measure positions of theses single electron in detector matrix~\cite{photos}. This way they literally obtained photography of orbitals: densities by averaging over positions of single electrons. Anyway, using Heisenberg principle as an excuse for ignoring questions about objective dynamics is no longer valid when increasing the scale, for example while asking question about local currents in a lattice: what is the probability distribution for electron jumping to neighboring parts of this lattice - getting a stochastic model for its dynamics (conductance). There are also examples of larger objects for which we should expect objective positions and so stochastic models, but in some situations their quantum description works surprisingly good, for example the nuclear shell model for baryons - MERW shows that such success of QM formalism does not disqualify stochastic description, in contrast it is also supported from this perspective, there is universality of quantum predictions.

\begin{figure}[t!]
    \centering
        \includegraphics{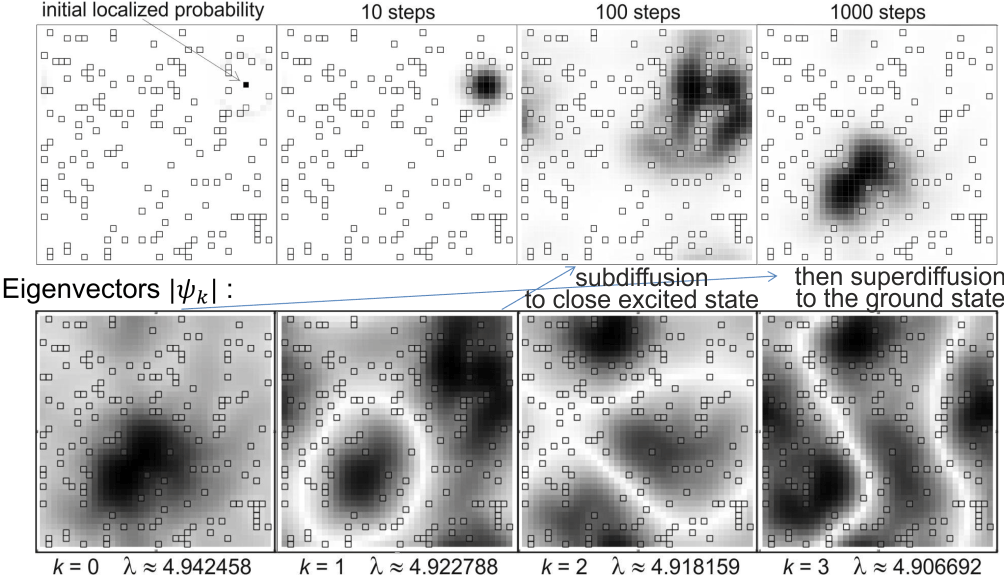}
        \caption{While MERW evolution leads to stationary probability distribution exactly as in the quantum ground state, the excited quantum states are also seen in this evolution. Specifically, the coefficient corresponding to $M\psi_i =\lambda_i \psi_i$ eigenfunction/eigenvector, drops $\approx (\lambda_i/\lambda_0)^t$ times during $t$ steps, where $\lambda_0$ is the dominant (largest) eigenvalue. As energy turns out to correspond to $-\lambda$, we get exponential weakening of contributions of excited states - in contrast to standard QM they are not stable, however, excited states in nature are also unstable. The top row of above diagram shows example of such evolution: as initial contribution of the ground state is very small, the evolution first gets close to the first excited state (subdiffusion), then it kind of tunnels to the ground state (superdiffusion). This simple model neglects conservation laws e.g. of energy, which might prevent the walker from going to the ground state - its perturbed averaged trajectory should instead smoothen toward some close excited probability distribution. Also for multiple repelling walkers (like electrons), they should choose densities of successive eigenstates here (like in Pauli exclusion principle).}
       \label{eig}
\end{figure}

\begin{figure}[t!]
    \centering
        \includegraphics{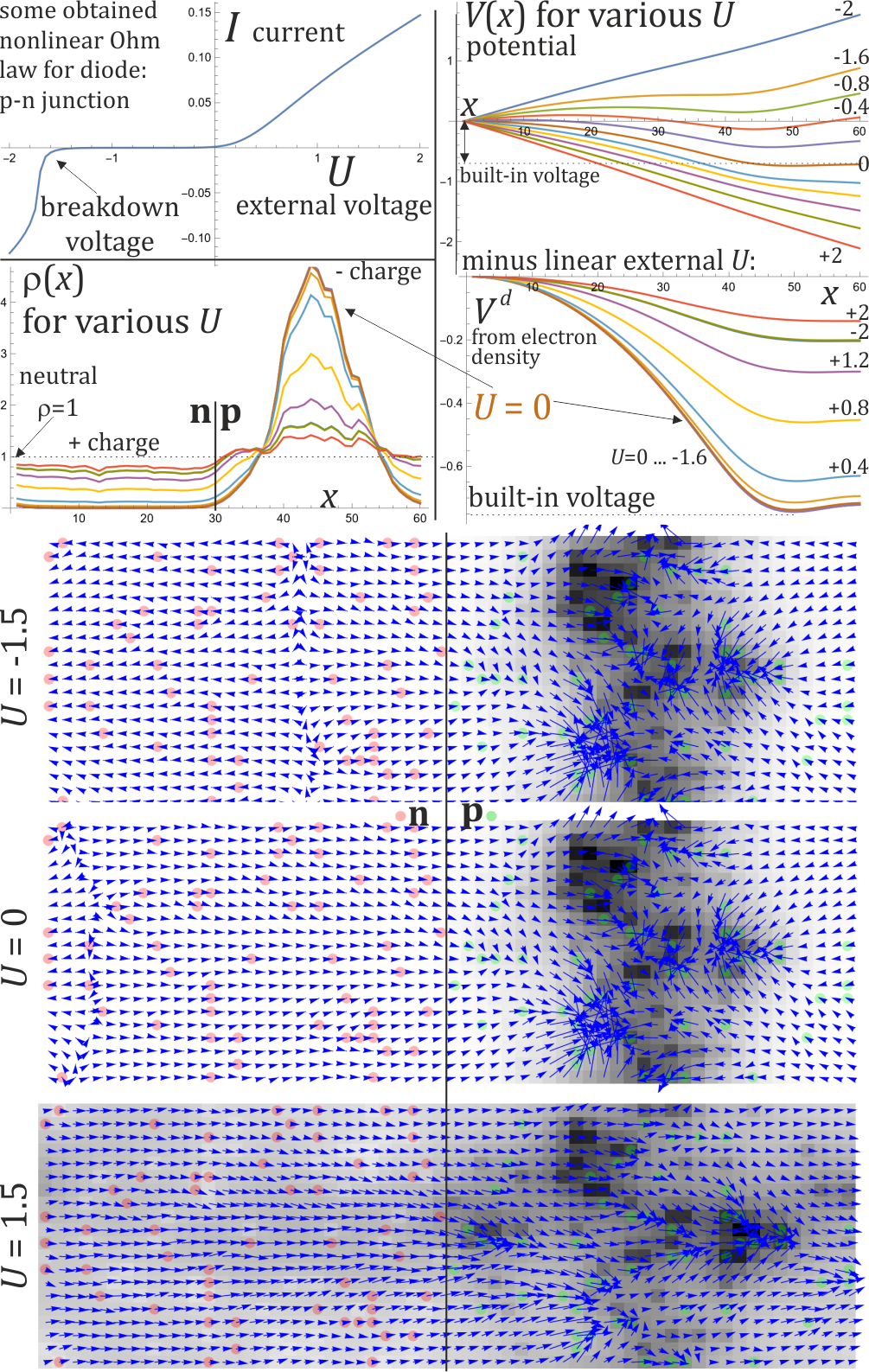}
        \caption{MERW model of conductance in p-n junction (diode)~(\cite{cond1}, simple simulator: \cite{cond}) for $60\times 30$ lattice with dopants: defects as nodes with lowered (p) or increased (n) potential. In GRW we would get nearly uniform electron density, leading to conductance for any external voltage as in (linear) Ohm law, incorrectly making it a conductor.
        In contrast, MERW has the same stationary probability distribution as QM - with Anderson localization preventing conductance (experimental using scanning tunneling microscope can be found in \cite{exp}), requiring some minimal breakdown voltage to start conductance in reverse bias, getting current-voltage dependence as in diode.}
       \label{cond}
\end{figure}

MERW allows to understand and repair this problem for example of seeing electron conductance as some statistical flow of charges - also where standard diffusion models had to give up, like defected lattice of semiconductors. The reason for this disagreement of standard stochastic models is that what looked as a natural choice for transition probabilities (like GRW) or stochastic propagator, often turns out only approximation of what is expected by statistical physics: entropy maximization. By repairing this approximation, MERW turns as close QM as we could expect from a diffusion model, like recreating equilibrium probability distribution exactly as the quantum ground state density. Also probability densities of excited states appear there as preferred, but can diffuse further (unless adding some constraint) like in Fig. \ref{eig}.

Hence MERW can be seen as quantum correction to diffusion models. However, this is still only diffusion, not a complete QM - it ignores interference, which requires e.g. some internal clock (de Broglie's $E=mc^2=\hbar\omega$, \href{https://en.wikipedia.org/wiki/Zitterbewegung}{zittebewegung}) of particle. Beside providing clear intuitions for looking problematic properties of QM (like squares in formalism leading to violation of Bell's inequalities), like in Fig. \ref{cond} such quantum corrections to diffusion can be also useful especially as practical approximations of extremely demanding complete quantum modeling of conductance: in semiconductor, microscopic scale, or single molecule electronics.

In the third Section there will explained MERW's analogue of measurement, especially for violation of Bell's inequalities. Fourth Section focuses on quantum computation - we will argue that Shor's algorithm also exploits the fact that we live in a spacetime, suggest a general approach for designing quantum algorithms. There is also discussed hypothetical approach for time-loop computers. Finally the last Section briefly discuses a possible ways to expand this simple but surprisingly successful effective model: just Boltzmann distribution among possible paths, into a more complete picture of physics, effectively described by quantum field theories - in perturbative approximation using ensemble of scenarios with varying number of particle: Feynman diagrams.\newpage

\section{Maximal Entropy Random Walk\\ as quantum corrections to diffusion}
Let us start with the common problem of choosing a random walk (as Markov process) on a graph defining the space of interest - which later will be chosen for example as a lattice, where we can introduce inhomogeneity (defects) like in Fig. \ref{merw}, or perform infinitesimal limit to get diffusion as continuous random walk. This section contains a condensed informal introduction, more complete description can be found as PhD Thesis of the author~\cite{myphd}, here is \href{https://community.wolfram.com/groups/-/m/t/2924355}{light introduction}.

\begin{figure}[t!]
    \centering
        \includegraphics{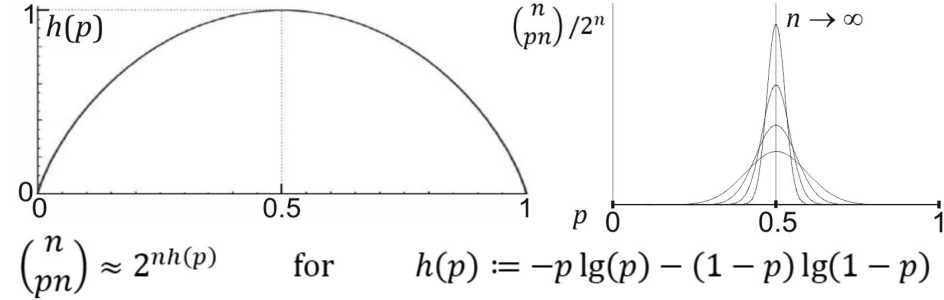}
        \caption{If among all 0/1 sequences of length $n$ we focus on subset of sequences with $p$ probability of '1', the size of this subset grows exponentially with $n\cdot h(p)$, where $h$ is Shannon entropy. In the $n\to \infty$ thermodynamical limit, the $p=1/2$ subset completely dominates all sequences - uniform probability distribution among all sequences has asymptotically $p=1/2$, what is a special case of Asymptotic Equipartition Property. Generally, entropy being such exponent leads to the Principle of maximum entropy: the safest choice of statistical parameters is the one maximizing entropy, in a random configuration we asymptotically should almost certainly get these statistical parameters.  }
       \label{entr}
\end{figure}

For simplicity assume here that the graph is indirected and defined by its (symmetric) adjacency matrix: $M_{ij}=M_{ji}=1$ if there is edge between vertex $i$ and $j$, 0 otherwise. From the perspective of physics, this adjacency matrix can be seen as simplified (zero potential) Bose-Hubbard Hamiltonian for a particle travelling between a set of sites connected as in this graph, jumping for $i$ to $j$ is first annihilation $a_i$ then creation $a_j^+$:
\be \mathcal{H}=-\sum_{\textrm{edge }(ij)} a_j^+ a_i \equiv - M \label{BH}\ee
\subsection{Standard random walk (GRW) and its suboptimality}
We would like to choose a stochastic matrix $S$ for this graph: $S_{ij}=\Pr(\gamma_{t+1}=j|\gamma_t=i)$, which is nonzero only for graph edges, outgoing probabilities for each vertex have to sum to 1:
\be 0\leq S_{ij}\leq M_{ij}\leq 1,\quad\qquad \forall_i \sum_j S_{ij}=1. \label{stoch} \ee
The standard way to choose random walk, referred as Generic Random Walk (GRW), is assigning equal probability to each outgoing edge, what for indirected graph leads to stationary probability distribution $(\sum_i \rho_i S_{ij}=\rho_j$)  with probability of vertex being proportional to its degree $d$:
\be S_{ij}^G=\frac{M_{ij}}{d_i}\qquad \rho^G_i=\frac{d_i}{\sum_j d_j}\qquad \textrm{for}\quad d_i=\sum_j M_{ij}\ee
Before commenting the above choice, let us remind the Principle of maximum entropy of Jaynes~\cite{jaynes}. Imagine a length $n$ sequence of '0' and '1', the number of such sequences is $2^n$. Now focus on subspace of possibilities with density $p\in[0,1]$ of value '1': with approximately $pn$ of '1'. Using Stirling formula $\left( k!\approx \sqrt{2\pi k}\,(k/e)^k\right)$ we can find asymptotic number of such combinations, plotted in Fig. \ref{entr}:
$${n \choose pn}\approx 2^{nh(p)}\quad\textrm{for} \quad h(p)=-p \lg(p)-(1-p) \lg(1-p)$$
\noindent being the Shannon entropy ($\lg\equiv \log_2$), which has single maximum $h(1/2)=1$. Hence splitting the set of all 0/1 sequence into disjoint subsets with $p$ density, asymptotically ($n\to \infty$) the $p=1/2$ uniform probability case will combinatorially completely dominate all the other subsets.

\begin{figure}[t!]
    \centering
        \includegraphics{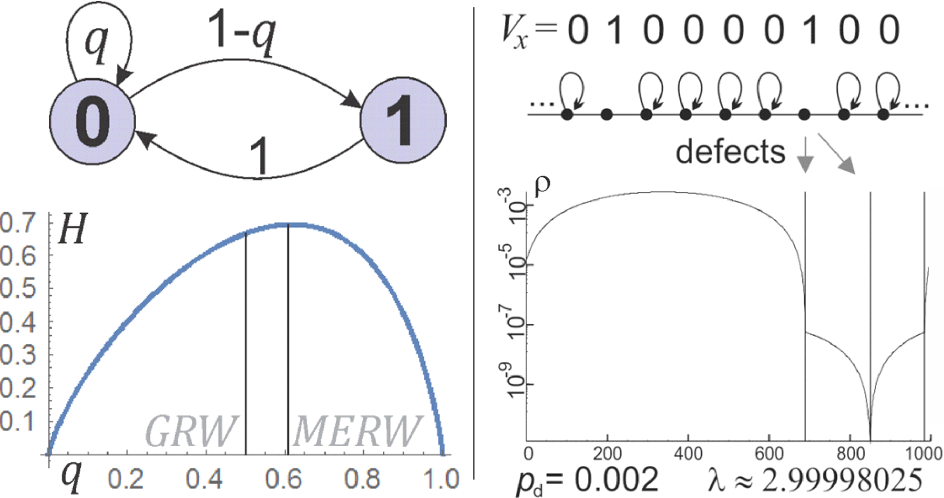}
        \caption{Left: Fibonacci coding as simple nontrivial example of suboptimality of GRW repaired by MERW. We have 0/1 sequences for which it is forbidden to use '11'. So after '0' we can use '0' or '1', but after '1' we have to use '0'. While seeing such sequence as random walk, the remaining question is to choose probability of '0' after '0' (parameter $q$). GRW suggests to choose $q=1/2$, but its entropy rate $H(S)$ is suboptimal: subset of sequences described by such parameter is asymptotically completely dominated by other sequences without '11'. In contrast, MERW chooses golden ratio $q=(\sqrt{5}-1)/2\approx 0.618$, maximizing $H(S)$ and properly describing statistics in the set of all sequences without '11', or in such single "typical" infinite sequence. Right: MERW on 1D defected lattice with cyclic boundary conditions: all vertices are connected with 2 neighbors, all but the marked defects have additional self-loop (can remain in the vertex). As stationary probability distribution of GRW is proportional to degree of vertex, we get 2/3 times lower density in defects. In contrast to such very weak localization property, the drawn density of MERW has very strong localization in the largest defect-free region (so called Lifshitz sphere), exactly as for the quantum ground state for this lattice. }
       \label{ex}
\end{figure}

Generally, like in the famous Boltzmann's formula, entropy is just (normalized) logarithm of the number of possibilities. Hence focusing on subset described by parameters (like density), maximizing entropy means focusing asymptotically on nearly all possibilities - contribution of subsets corresponding to suboptimal parameters asymptotically vanishes in exponential way with the size of the system. It can be summarized in the \textbf{Principle of maximum entropy: probability distribution which best represents the current knowledge is the one with largest entropy}. Without additional knowledge, entropy is maximized for uniform probability distribution on a given set. Assigning energy to objects/possibilities and fixing total energy, we get Boltzmann distribution instead. These two distributions are the base of statistical physics.\\

Returning to random walk on a graph, GRW clearly maximizes entropy for every vertex - is kind of local maximization. The question is if it maximizes average entropy per step: averaged over probability distribution of being in a given vertex. This measure is also called entropy rate:
\be H(S)=\sum_i \rho_i \sum_j S_{ij} \lg(1/S_{ij})\ee
\noindent It turns out to be equal to normalized entropy in the space of sequences generated by such Markov process $S$:
\be H(S)=\lim_{n\to\infty} \frac{1}{n}\sum_{\gamma =\gamma_0\ldots\gamma_n} \Pr(\gamma)\lg(1/\Pr(\gamma))\label{rate}\ee
$$ \textrm{where}\qquad \Pr(\gamma_0\ldots\gamma_n)=\rho_{\gamma_0} S_{\gamma_0 \gamma_1}\ldots S_{\gamma_{n-1}\gamma_n}$$
\noindent is probability of obtaining sequence $\gamma$.

By Maximal Entropy Random Walk (MERW) we will refer to the choice of $S$ matrix which maximizes $H(S)$ over all random walks on a given graph: fulfilling conditions (\ref{stoch}). As this maximization involves corresponding stationary probability distribution: dominant eigenvector of $S$ matrix to eigenvalue 1: $\rho S=\rho$, for maximization it is more convenient to use formula (\ref{rate}), which reaches maximum for uniform probability distribution among (infinite) paths generated by a given Markov process.

In many cases GRW already maximizes $H(S)$ making it equal with MERW, for example for regular graphs (all vertices have the same degree), like regular lattice and so standard diffusion in empty homogeneous space obtained as continuous limit of the lattice. The simplest example of nonoptimality of GRW is Fibonacci coding case, presented in Fig. \ref{ex}. More physical examples are defected lattices, for example representing a semiconductor, or its continuous limit: diffusion in inhomogeneous space. In contrast to standard diffusion which leads to nearly uniform stationary probability distribution, MERW leads to very strong localization properties - exactly as the quantum ground state (for e.g. Bose-Hubbard or Schr{\"o}dinger Hamiltonian), what we would from QM consideration and for example prevents semiconductor from being a conductor by prisoning electrons in entopic wells as mentioned in Fig. \ref{merw}.

The GRW nearly uniform stationary probability distribution can be seen as maximizing entropy in spatial "static 3D" picture like in Fig. \ref{born}: for a fixed time cut of spacetime, leading e.g. to unform density in $[0,1]$ potential well there. In contrast, MERW philosophy maximizes entropy in 4D spacetime picture: where particles become their trajectories, leading to $\rho(x)\propto \sin^2(\pi x)$ there, exactly as predicted also by QM.

\begin{figure}[t!]
    \centering
        \includegraphics{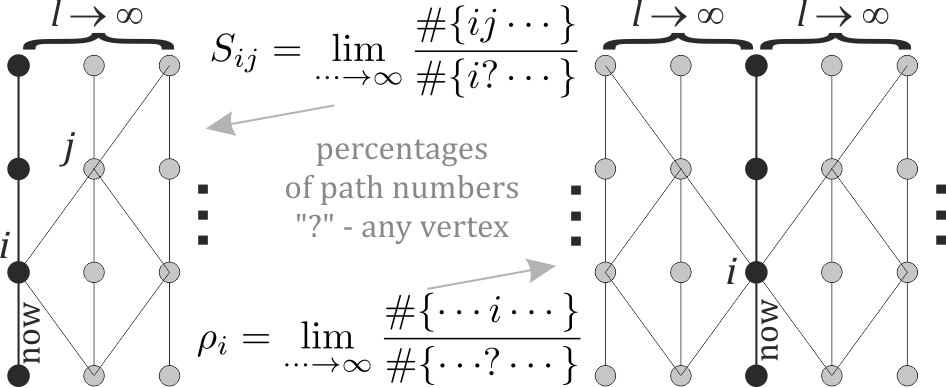}
        \caption{Deriving MERW formulas for simple four-vertex graph: term in its $4\times 4$ adjacency matrix $M_{ij}$ is 1 iff $|i-j|=1$. Left: terms of stochastic matrix $S_{ij}$ can be found by considering all length $t$ paths starting in $i$ and calculating what percentage of them makes the first step to $j$. For $t=1$ we GRW this way, generally we can also consider $\textrm{GRW}_t$ family for $t\in\mathbb{N}^+$, finally in $t\to\infty$ limit we obtain MERW stochastic matrix. Right: MERW stationary density can be found by considering all length $2t+1$ trajectories: $\rho_i$ is percentage of those having $x$ in the center in $t\to \infty$ limit is $\rho_x$. These formulas come from assumption of unique dominant eigenvalue of adjacency matrix: $M^t \to \lambda^t \psi \psi^T$.}
       \label{merwform}
\end{figure}
\subsection{MERW formulas and Born rule}
While GRW assumes uniform probability distribution among outgoing edges: paths of length one, let as analogously define $\textrm{GRW}_k$ to assume uniform probability distribution among length $t$ paths ($\textrm{GRW}\equiv\textrm{GRW}_1$). The number of length $t$ paths from vertex $i$ to $k$ for which the first step is to $j$ is $M_{ij}(M^{t-1})_{jk}$, hence the stochastic matrix of $\textrm{GRW}_t$ is $S_{ij}\propto M_{ij} \sum_k (M^{t-1})_{jk}$.

MERW assumes uniform probability distribution among possible infinite paths, what allows to see it as $t\to\infty$ limit of $\textrm{GRW}_t$ like in Fig. \ref{merwform}. To derive formula for $t\to\infty$ limit, for simplicity let us assume that our indirected graph ($M=M^T$) is connected and aperiodic, where the Frobenius-Perron theorem says that $M$ has non-degenerated (single) dominant eigenvalue $\lambda > 0$ and its corresponding eigenvector (left and right are equal for symmetric $M$) has real nonnegative coordinates:
$$ \psi M =M\psi = \lambda \psi \quad\textrm{maximizing }\lambda\in \mathbb{R}^+,\quad \psi_i\in \mathbb{R}^+ $$
Non-degenerated dominant eigenvalue makes that in the $t\to \infty$ limit we have $M^t \to \lambda^t \psi \psi^T $ (or $\lambda^t|\psi\rangle \langle \psi|$ in bra-ket formalism), getting MERW $S_{ij}\propto M_{ij} \psi_j$ as limit of $\textrm{GRW}_t$. For normalization, as $\sum_j M_{ij} \psi_j =\lambda \psi_j$, we get the final \textbf{formula for MERW} stochastic matrix:
\be S_{ij}=\frac{M_{ij}}{\lambda} \frac{\psi_j}{\psi_i} \qquad\quad \rho_i \propto \psi_i^2 \label{merwform} \ee
\noindent where the above formula for stationary probability distribution, $\rho_i = \psi_i^2/\sum_j \psi_j^2$ after probabilistic normalization, can be easily verified:
$$\sum_i \psi_i^2 \cdot \frac{M_{ij}}{\lambda} \frac{\psi_j}{\psi_i} = \sum_i \psi_i M_{ij} \frac{\psi_j}{\lambda} = \lambda \psi_j \frac{\psi_j}{\lambda} = \psi_j^2$$
The above derivation of MERW stochastic formula has used ensemble of infinite half-paths going forward in time, with (normalized) $\psi$ describing probability distribution at the beginning of such ensemble. To analogously derive the formula for its stationary probability distribution, we can fix a position $i$ and use ensemble of infinite half-paths toward both past in future. Using bra-ket formalism, both derivation can be informally written as:
$$ |i\rangle \langle i | |i\rangle \langle j |M^{t-1} \to\ \propto \lambda^{t-1} \psi_j \quad \textrm{imples}\quad S_{ij}\propto M_{ij} \psi_j $$
\be M^t |i\rangle \langle i | M^t \to\ \propto \lambda^{2t} \psi_i ^2 \quad \textrm{imples}\quad \rho_i \propto \psi_i ^2 \label{pathen}\ee
This way we get a natural intuition for $\rho_i \propto \psi_i^2$ containing \textbf{Born rule} as in Fig. \ref{born}, \ref{born1}: the quantum amplitude $\psi$ describes situation at the beginning of past or future half-spacetime (usually equal). If we want to measure a position or some observable in a given moment, we need to "draw" this random value from both past and future directions, getting final probability being product of both original (identical) probabilities, getting squares known from the quantum formalism, which as we know for example lead to violation of Bell inequalities wrongly expected by our natural "evolving 3D" intuition.

\subsection{Boltzmann path distribution}
Observe that taking a power of MERW stochastic matrix (\ref{merwform}), or calculating probability distribution of a path $\gamma$, the intermediate $\psi_j/\psi_i$ terms cancel, getting:
$$(S^t)_{ij}=\frac{(M^t)_{ij}}{\lambda^t} \frac{\psi_j}{\psi_i}$$
\be \Pr(\gamma_0\ldots\gamma_t)=\frac{\psi_{\gamma_0}\ M_{\gamma_0\gamma_1}\ldots M_{\gamma_{t-1}\gamma_t}\  \psi_{\gamma_t}}{\lambda^t} \label{dist}\ee
This way we get another "local" equivalent condition for MERW (written in Fig. \ref{merw}): for all two vertices, each path of given length between them is equally probable.

In statistical physics we emphasize some possibilities by introducing energy, going from uniform to Boltzmann distribution $p_i\propto e^{-\beta E_i}$, which is  obtained from the Jaynes Principle of Maximum entropy by maximizing entropy under constraint of fixed average energy.

To take Boltzmann distribution to the MERW philosophy we need first to define energy of paths. A simple way is through choosing energy (potential) corresponding to each edge: $V_{ij}$, then define energy of a path as sum over all its edges:
$$\textrm{energy of path }(\gamma_0\ldots\gamma_t)\quad \textrm{ is }\quad V_{\gamma_0 \gamma_i}+\ldots+V_{\gamma_{t-1} \gamma_t}$$
In equation (\ref{dist}) we can change from uniform to such Boltzmann distribution among paths by just using more general $M$ matrix: still real nonnegative, but carrying weights corresponding to related potential:
\be M_{ij}=A_{ij}\,e^{-\beta V_{ij}}\qquad \textrm{for}\quad A_{ij}\in\{0,1\} \label{boltz}\ee
\noindent being the adjacency matrix.

To use vertex potential $V_i$ instead, we can take e.g. $ V_{ij} = \frac{1}{2} \left(V_i + V_j \right).$
\subsection{Lattice and continuous limit to Schr{\"o}dinger equiation}

As discussed, adjacency matrix can be seen as minus simplified (without potential) Bose-Hubbard Hamiltonian (\ref{BH}). Lattices are basic graphs used in physics, continuous situation can be realized as infinitesimal limit of a lattice. Let us start with 1D lattice from Fig. \ref{ex}: MERW with potential barriers realized using self-loops (1 at diagonal of adjacency matrix, possibility to remain in the vertex), here removed in defects: $V_x=0$ if vertex $x$ contains self-loop, 1 otherwise.

While for GRW stationary probability distribution is proportional to degree of a vertex, getting very weak localization, for MERW we first need to find the dominant eigenvector of adjacency matrix ($\lambda\psi=M\psi$):
$$(\lambda \psi)_x = \psi_{x-1} + (1-V_x) \psi_x +\psi_{x+1} \qquad / -3\psi_x,\ \cdot -1$$
\be E \psi_x = -(\psi_{x-1}-2\psi_x + \psi_{x+1}) +V_x \psi_x \label{s1}\ee
\noindent where maximization of $\lambda$ has became minimization of energy $E:= 3- \lambda$ due to change of sign.

The $(\psi_{x-1}-2\psi_x + \psi_{x+1})$ term is discrete Laplacian, making formula (\ref{s1}) discrete analogue of stationary Schr{\"o}dinger equation. Hence MERW predicts going to exactly the same stationary probability distribution $\rho_x\propto \psi_x^2$ as predicted by quantum mechanics here - with very strong localization properties, for example preventing semiconductor from being a conductor.

To get the standard 1D continuous Schr{\"o}dinger equation, let us take a regular lattice with $\epsilon$ time step and $\delta$ lattice constant. To consider real potential $V$, we can assume Boltzmann distribution among paths as in (\ref{boltz}). The $\lambda_{\epsilon}\psi_x =(M_{\epsilon}\psi)_x$ eigenequation becomes for example:
$$\lambda_{\epsilon}\psi_x =e^{-\beta\epsilon V_{x-1}}\psi_{x-1}+e^{-\beta\epsilon V_{x}}\psi_{x}+e^{-\beta\epsilon V_{x+1}}\psi_{x+1}$$

As we are interested in $\epsilon, \delta\to 0$ limit, let us use approximations: $e^{-\beta\epsilon V}\approx 1-\beta\epsilon V$ and $V_{x-1}\approx V_x \approx V_{x+1}$. After simple transformations and multiplying by -1 as previously (to change from maximization of $\lambda$ to minimization of $E$), we get:
$$\frac{3-\lambda_\epsilon}{3\beta\epsilon}\psi_x \approx -\frac{1}{3\beta} \frac{\psi_{x-1}+\psi_{x}+\psi_{x+1}}{\epsilon} + V_x \psi_x$$

Defining energy and choosing relation between time and space steps: with square required for getting from discrete to continuous random walk (as standard deviation grows with square root of time):
$$E_\epsilon = \frac{3-\lambda_\epsilon}{3\beta\epsilon}\qquad\qquad \epsilon =\frac{\delta^2}{3\alpha}$$
\noindent for some $\alpha$ parameter, in the $\delta\to 0$ limit we get standard 1D stationary Schr{\"o}dinger equation:
\be E\Psi =\left( -\frac{\alpha}{\beta}\Delta + V\right)\Psi=\left( -\frac{\hbar^2}{2m}\Delta + V\right)\Psi \ee
\noindent assuming $\alpha/\beta = \hbar^2/2m$. Considering time dependant situation and comparing the continuity equations~\cite{myphd}, suggests to choose: $\alpha = \hbar / 2m,\ \beta = 1/\hbar$. For such generalized MERW there are also appearing other QM properties like Heisenberg principle.

The nature and values of such constants describing "fundamental noise" are crucial but not well understood. A natural source might be intrinsic periodic process of particles, so called zitterbewegung or de Broglie's clock $E=\hbar \omega$, which has been directly observed for electrons~\cite{hest,elclock}. The MERW behavior also sees excited states, as presented in Fig. \ref{eig}, suggesting that random thermodynamical distortion of a classical trajectory should make it smoothen toward probability cloud of close (overlapping) potentially excited quantum state.

What is surprising in the above lattice derivations is that Laplacian, which in standard QM is related with momentum, here appears only from spatial structure of the lattice as these corrected diffusion models do not consider velocity of particle. To add kinetic energy into considerations, we could perform MERW in phase space: (space, velocity), like in the Langevin equation, however, it becomes much more complicated. \\

\begin{figure}[t!]
    \centering
        \includegraphics{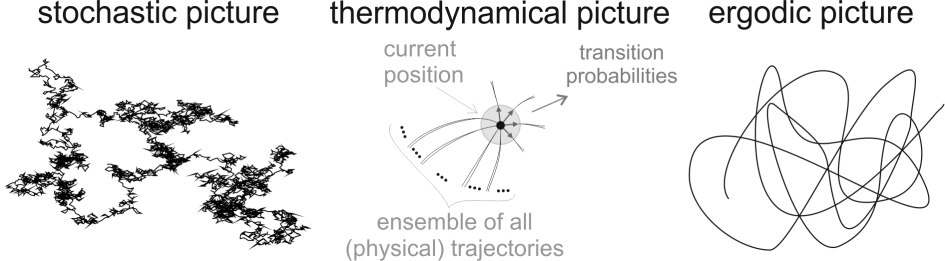}
        \caption{Three philosophies for probabilistic modelling of dynamics. Left: stochastic picture where evolution is imagined as succeeding random choices, with arbitrarily chosen probabilities e.g. locally maximizing entropy (GRW), for which it has nearly no localization property. Right: ergodic picture usually assuming a complex deterministic evolution (neglecting e.g. thermodynamical fluctuations), density is obtained by averaging over e.g. a single trajectory. Center: the discussed MERW-like philosophy based the principle of maximum entropy required by thermodynamical model, using Boltzmann distribution among possible trajectories, leading to quantum-like densities. Local transition probabilities (stochastic propagator) are calculated from ensemble of trajectories. The object does not directly use these probabilities (nonlocal: depending on the entire system), just performs some complex dynamics - only we use probabilities of this thermodynamical model to predict the safest evolution of our knowledge.}
       \label{gen}
\end{figure}

The fact that Boltzmann distribution among paths leads to quantum thermodynamical predictions is not surprising as this MERW philosophy seems close to Feynman's Euclidean path integrals (EPI), however, there are some essential differences:
\begin{itemize}
\item Philosophy: EPI starts with assuming the axioms of QM, then performs "Wick rotation" to imaginary time - both having questionable clarity. In contrast, MERW just repairs diffusion: accordingly to the principle of maximum entropy, repairing known disagreements with reality of standard diffusion.
\item Formula: standard EPI propagator lacks stochastic normalization, especially the crucial $\psi_j/\psi_i$ term, which modifies the behavior with position in a very nonlocal way (dependent on the entire system).
\item Statistics: standard EPI assumes Boltzmann distribution among paths in a given time period, like in the $\textrm{GRW}_t$ philosophy - conditioning the behavior on this arbitrarily chosen time period. In contrast, MERW uses ensemble of paths infinite in both time directions.
\item Complexity: EPI starts with the continuous case, which path integration is mathematically problematic. In contrast, MERW philosophy starts with well understood discrete case.
\end{itemize}
Mathematically closer to MERW is Zambrini's Euclidean quantum mechanics~\cite{zambrini}, but like for Nelson's stochastic quantum mechanics~\cite{nelson}, the motivation is fitting the expected behavior of QM, instead of MERW's just concluding from required fundamental mathematical principle: of entropy maximization.

\section{Measurement and Bell inequalities} \label{bellsect}
While our intuition of living in 3D space evolving in time requires satisfaction of Bell-like inequalities, they can be violated in real world or QM - let us understand it from perspective of living in 4D spacetime instead: where the basic objects are trajectories and we should consider their ensembles like in Euclidean path integrals or MERW.

For simplicity let us focus on Mermin's~\cite{mermin} Bell-like inequality for three binary variables $A,B,C\in\{0,1\}$:
\be \textrm{Pr}(A=B)+\textrm{Pr}(A=C)+\textrm{Pr}(B=C)\geq 1 \label{mermin} \ee
It can be intuitively explained that \textbf{tossing three coins, at least two of them give the same outcome}. More formally, choosing any probability distribution among their $2^3=8$ possibilities $\left(\sum_{ABC=0}^1 p_{ABC}=1\right)$, each of three equalities correspond to 4 out of 8 possibilities - as shown in Fig. \ref{bell}, summing $\textrm{Pr}(A=B)+\textrm{Pr}(A=C)+\textrm{Pr}(B=C)$ we get $\left(\sum_{ABC=0}^1 p_{ABC}\right) +2p_{000}+2p_{111}\geq 1$.

While it seems impossible for this inequality to be false, it is somehow violated in quantum formalism. For this purpose, it is crucial that we measure only 2 out of 3 variables, otherwise we would operate on $\{p_{ABC}\}$ probability distribution - which satisfies the (\ref{mermin}) inequality.

So we measure 2 out of 3 variables - each outcome represents two possibilities for the unmeasured variable, e.g. $AB?$ outcome represents $\{AB0,AB1\}$ set. To violate the inequality we need something nonintuitive, like Born rules characteristic for QM and MERW:

\begin{itemize}
  \item \textbf{Kolmogorov 3rd axiom}: probability of alternative of disjoint events is sum of individual probabilities: $p_{AB?}=p_{AB0}+p_{AB1}$, leading to the inequality (\ref{mermin}).
  \item \textbf{Born rule}: probability of alternative of disjoint events is proportional to \textbf{square} of sum of their amplitudes: $p_{AB?}\propto (\psi_{AB0}+\psi_{AB1})^2$.
\end{itemize}

As in Fig. \ref{bell}, this Born rule assumption allows to violate inequality (\ref{mermin}): for example taking $\psi_{000}=\psi_{111}=0$, $\psi_{001}=\psi_{010}=\psi_{011}=\psi_{100}=\psi_{101}=\psi_{110}>0$ we get $\textrm{Pr}(A=B)=\textrm{Pr}(A=C)=\textrm{Pr}(B=C)=1/5$ and so violation of the inequality to $3/5<1$.

Assuming as in Euclidean path integrals or MERW: uniform probability distribution among paths, we got the squares like in Born rules by multiplying amplitudes from both time directions. To formalize it we need to define MERW measurement, which in QM is destructive process: transforms usually continuous initial state into a discrete set of possibilities: eigenvectors of measurement operator.

To understand destructiveness of measurements, adapt it to a simple model like MERW, let us look at \textbf{Stern-Gerlach experiment} which is  used as idealization of measurement: it applies strong magnetic field to transform initially continuous space of spin directions into one of two possibilities: parallel or anti-parallel alignment (only these two do not have Larmor precession hence minimize kinetic energy), which can be later separated using field gradient.

We can imagine that after aligning in strong magnetic field, spin can no longer flip during this flight in strong magnetic field - it leads to idealized condition which can be adapted for MERW and turns out sufficient: \textbf{during measurement its outcome cannot change}.

\begin{figure}[t!]
    \centering
        \includegraphics{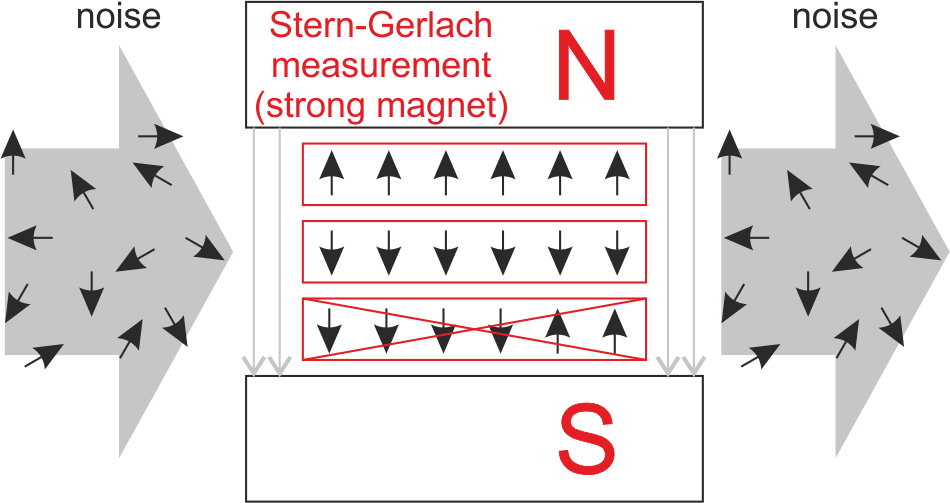}
        \caption{Measurement is a destructive process. We need to extract idealization of its influence on the system to take it to ensemble of paths considered here. Stern-Gerlach experiment is seen as idealization of measurement: with strong magnetic field enforcing initially random spin direction to choose between aligned or anti-aligned direction. This choice cannot be changed during the measurement: strong magnetic field does not allow to flip spin after it was already chosen. Such rule can be taken to ensemble of paths and turns out sufficient to get Born rules: we will \textbf{restrict ensemble to paths not changing outcome during measurement}.}
        \label{sterngerlach}
\end{figure}

\subsection{Realization of Bell violation example}
\begin{figure}[t!]
    \centering
        \includegraphics{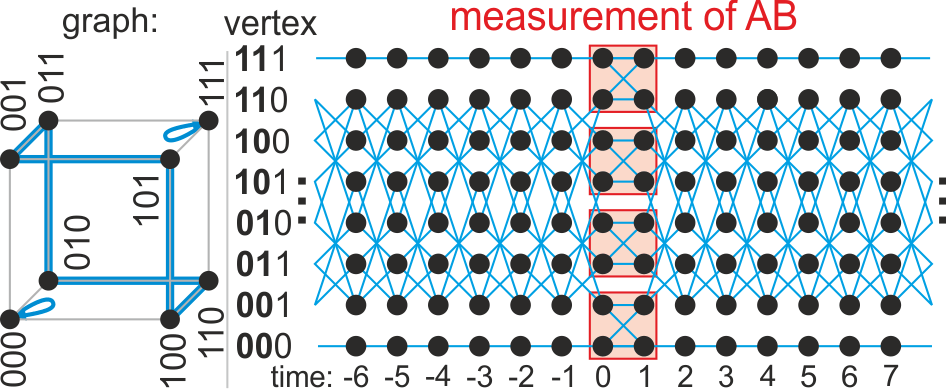}
        \caption{Left: graph we are considering, allowing for steps using the blue edges. Right: considered trajectories among which we assume uniform probability distribution (MERW) as analogue of Stern-Gerlach situation from Fig. \ref{sterngerlach}, noise before and after measurement corresponds to random walk. Each column represents 8 possible vertices from assumed graph (cube), column's number represents time moment. All but the measurement moment we assume allowed paths accordingly to the marked (blue) edges of the cube (can be also used simpler, e.g. width 3 Ising lattice with interaction forbidding $\downarrow\downarrow\downarrow$ and $\uparrow\uparrow\uparrow$). This behavior is interrupted by measurement of $AB$ coordinates in 0-1 time step (marked red) - single transition enforcing to remain outcome value: using only edges inside marked squares. Probability of given measurement outcome (square) is the number of paths going through its square, divided by the total number of paths. }
        \label{bellmerw}
\end{figure}

Let us take "during measurement its outcome cannot change" rule to MERW like in Figure. \ref{bellmerw}. In all but time 0-1 step there are allowed steps accordingly to the assumed graph: blue edges in cube on the left (can be also used simpler, e.g. width 3 Ising lattice with interaction forbidding $\downarrow\downarrow\downarrow$ and $\uparrow\uparrow\uparrow$, we measure 2 out of 3 spins). In the remaining time 0-1 step we measure $AB$: first two out of three variables. This step is governed by the measurement rule: it cannot change the measured coordinates. However, it can change the third (unmeasured) coordinate - which is volatile in this measurement, otherwise inequality (\ref{mermin}) would be satisfied.

For this spacetime diagram presented in the right part of Fig. \ref{bellmerw}, let us assume MERW rule that all possible paths (using blue edges) are equally probable - asking what percentage of them goes through the four boxes corresponding to measurement outcomes, we will correspondingly get $1/10,\ 4/10,\ 4/10,\ 1/10$ probabilities, which violate the inequality.

Specifically, let us calculate the number of past paths from time $t=-l$ to $t=0$ in Fig. \ref{bellmerw}. For the 000 and 111 final vertices there is only a single such path. For the remaining 6 vertices there are 2 possibilities for each time step, hence there are $2^l$ paths ending in each of these vertices. Analogously for future paths: from time $t=1$ to $t=l+1$, their number is 1 for the 000 and 111, and $2^l$ for the remaining 6 vertices.

Now let us count bidirectional paths: from $t=-l$ to $t=l+1$ going through each of 4 measurement boxes. For the top and bottom box, corresponding to measuring 00 or 11 of $AB$ coordinates, the number of paths is $(2^l+1)^2$. For the remaining two boxes, the number of paths is $(2^l+2^l)^2$. Asymptotically ($l\to\infty$) we get:
$$\textrm{Pr}(A=0,B=0)=\lim_{l\to\infty} \frac{(2^l+1)^2}{2(2^l+1)^2+2(2^l+2^l)^2}=\frac{1}{10} $$
and analogously for the remaining three measurement outcomes, getting probability of going through each of the four boxes being correspondingly: $1/10,\ 4/10,\ 4/10,\ 1/10$.

Hence, in this scenario $\textrm{Pr}(A=B)=2/10$ when measuring the first two coordinates. Analogously measuring the remaining two pairs (different grouping into pairs of vertices), we get violation of the inequality:
$$\textrm{Pr}(A=B)+\textrm{Pr}(A=C)+\textrm{Pr}(B=C)=6/10<1$$
\subsection{Born rules in general MERW measurement}
To generalize this Born rule to adjacency matrix $M$, assume that measurement chooses between disjoint subsets of possibilities (4 red squares in Fig. \ref{bellmerw}), splitting the set of vertices $X$ into disjoint subsets (components) distinguished by the measurement:
$$X=\bigsqcup_i X_i\qquad\textrm{and we need}\quad P(X_i)\propto \left(\sum_{j\in X_i} \psi_j\right)^2$$

Using the rule that \textbf{two neighboring steps are in the same component} during measurement as before:\\

\textbf{Definition:}  \emph{MERW measurement} in time $t=0$ for split $X=\bigsqcup_i X_i$ modifies the original uniform ensemble among paths $(\gamma_t)_{t\in\mathbb{Z}}$, to all paths satisfying: $\exists_i\ \gamma_0, \gamma_1 \in X_i$ (beside usual $M_{\gamma_t \gamma_{t+1}}=1$ for $t\neq 0$).\\

The number of paths from $t=-l$ to $t=l+1$ going through $X_i$ in time 0 and 1 is:
\begin{small}$$\sum_{ab} \sum_{jk\in X_i} (M^l)_{aj} (M^l)_{kb}\to \lambda^{2l} \sum_{jk\in X_i} \psi_j \psi_k =\lambda^{2l}\left(\sum_{j\in X_i} \psi_j\right)^2$$\end{small}
\noindent where $\psi$, $\lambda$ are dominant eigengenvector/eigenvalue ($M\psi=\lambda\psi$, assume it is unique), asymptotically $(l\to\infty)$ getting general Born rule:
$ \textrm{Pr}(X_i)\propto \left(\sum_{j\in X_i} \psi_j\right)^2$.

\section{Four dimensional understanding\\ of quantum computation}
While violation of Bell inequalities is rather only an interesting fact regarding consequences of QM, much deeper and applied exploitation of quantum strangeness is proposed for quantum computers, especially the Shor's algorithm~\cite{shor} shifting the factorization problem from exponential to polynomial complexity. This believed exponential classical cost is crucial for safeness of widely used asymmetric cryptography like RSA, which could be endangered by quantum computers if overcoming technical (or deeper?) difficulties of their implementation.

Such possibility of shifting from classical exponential to quantum polynomial complexity suggests some computational superiority coming with the nonintuitive properties of quantum mechanics - understanding of which might help us designing new quantum algorithms, especially to understand the question of existence of polynomial quantum algorithms for NP-complete problems, for which positive answer could, among others, endanger all kind of used cryptography.

\begin{figure}[!t]
    \centering
        \includegraphics{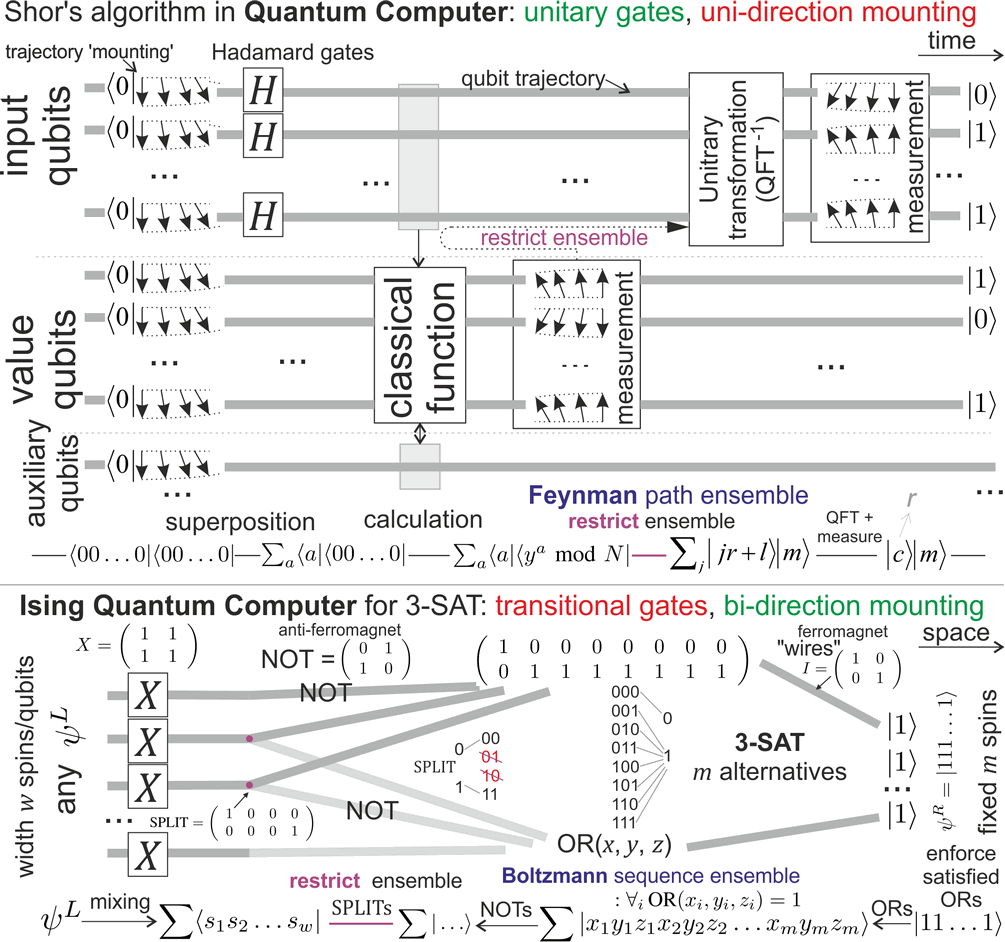}
        \caption{\textbf{Top}: Schematic diagram of quantum subroutine of Shor's algorithm~\cite{shor} for finding prime factors of natural number $N$. For a random natural number $y<N$, it searches for period $r$ of $f(a)=y^a \textrm{ mod }N$. This period can be concluded from measurement of value $c$ after Quantum Fourier Transform ($\textrm{QFT}^{-1}$) and with some large probability $(O(1))$ allows to find a nontrivial factor of $N$. The Hadamard gates produce state being superposition of all possible values of $a$. Then classical function $f(a)$ is applied, getting superposition of $|a\rangle |f(a)\rangle$. Due to necessary reversibility of applied operations, this calculation of $f(a)$ requires use of auxiliary qubits, initially prepared as $|0\rangle$. Now measuring the value of $f(a)$ returns some random value $m$, and restricts the original superposition to only $a$ fulfilling $f(a)=m$. Mathematics ensures that $\{a:f(a)=m\}$ set has to be periodic here $(y^r \equiv 1 \mod N)$, this period $r$ is concluded from the value of Fourier Transform ($\textrm{QFT}^{-1}$). Seeing the above process as a situation in 4D spacetime, qubits become trajectories, state preparation mounts their values (chosen) in the past (beginning), measurement mounts their values (random) in the future (end). Superiority of this quantum subroutine comes from future-past propagation of information (tension) by restricting the original ensemble in the first measurement. \textbf{Bottom}: NP complete problem (3SAT) approach with Ising model~\cite{ising2019} - restring Boltzmann path ensemble to corresponding to solutions of given instance of problem. It brings fundamental question if Boltzmann (or Feynman in quantum computers) ensemble of e.g. $2^{1000}$ paths is more than an idealization? }
       \label{fqc}
\end{figure}

One characteristic property of quantum algorithmics is the requirement to use only reversible operations (gates) as quantum evolution is unitary. Observe that $(x,y,z)\to (x,y,z \textrm{ XOR }g(x,y))$ is its own reverse and allows to realize any boolean function like AND, OR and XOR if using prepared auxiliary bit $z=0$. While we could classically reverse such gates and their sequences realizing some function, the requirement of a large number of prepared auxiliary bits prevents such use of reversible operations to actually reverse a difficult function, like the discrete logarithm - it would require fixing on both ends of the process: of final values of the function and initial values of the auxiliary bits.

Hence the question is if we could influence some complex (reversible/time symmetric) computational process on its both ends (initialization and output/measurement) in order to obtain a somehow more superior computation capabilities, e.g. shifting some problem from exponential to polynomial complexity? We could fix a system of rubber bands on its both ends, like for anyons forming braids in Kitaev's hypothetical topological quantum computers~\cite{top}. However, it seems technically difficult to realize logic gates on such rubber bands in 3D. Even if we could realize basic gates for them, minimizing the tension of such rubber band system might be physically very difficult to stabilize for solving our computational problem, especially that for hard problems the number of local energy minima grows exponentially with problem size~\cite{pnp}. Another way to fix values in both boundaries is replacing Feynman with Boltzmann path ensemble: going to Ising model as in bottom of Fig. \ref{fqc} - bringing question if such ensemble (also Feynman) of e.g. $2^{1000}$ paths is more than an idealization?

The situation seems more optimistic if this "rubber band setting" is in 4D spacetime: is a system of trajectories of some qubit carriers. One reason is that realizing logic gates is simpler in 4D than in 3D thanks to more freedom. More importantly, optimization of such system to solve our problem is no longer a continuous process, but from perspective of action optimization formulation of Lagrangian mechanics: we can imagine that  nature has already solved the problems we are planning to ask.

Figure \ref{fqc} contains such schematic picture for Shor's algorithm. Fixing situation in the past is easy: just prepare the qubits in some chosen states. Additionally, quantum measurement gives some possibility to affect the system also from the future direction: in case of Shor's algorithm it restricts the original ensemble to only possibilities having the same (randomly chosen) value of calculated function. As emphasized in this diagram, the consequence of this restriction (tension) seems to propagate backward in time here, like in Wheeler's or delayed choice quantum erasure QM experiments, or in action optimizing formulation of Lagrangian mechanics.

Hence the suggested general approach to exploit the quantum superiority e.g. to search for polynomial algorithm for some NP-complete problem is:
\begin{enumerate}
  \item Use Hadamard gates to get superposition of exponentially large set of possibilities, for example of all inputs to the problem among which we search for the satisfying one,
  \item Perform some chosen classical function on these inputs, getting superposition like $\sum_a |a\rangle |f(a)\rangle$,
  \item Measure value of this function, restricting the ensemble to $\sum_{a:f(a)=m} |a\rangle |m\rangle$,
  \item Ask a question about this final ensemble, for example about its periodicity using QFT. Another basic question we can realize is if the size of resulting superposition is larger than one, what can be done by first producing multiple copies of bits of the inputs (e.g. using $(x,y)\to (x, y\textrm{ XOR }x)$ for auxiliary $y=|0\rangle$), then measuring them: values of their measurements will vary iff the superposition contains more than one possibility.
\end{enumerate}

It seems tempting to implement with quantum gates e.g. verifier for an instance of NP-complete problem, and try to restrict ensemble to inputs satisfying this verifier, but in standard approach such restriction does not seem realisable. However, if being able to realize CPT analogue of state preparation, it might influence measurement outcomes, allowing to solve NP problems with quantum computers. Such hypothetical possibility is suggested in the next section using e.g. ring laser, which might be able to intuitively pull photons of chosen polarity from measured objects, hopefully changing probability distribution of measurement outcomes.

\section{Hypothetical negative photon pressure\\ and its potential applications} \label{slaser}
\begin{figure}[b!]
    \centering
        \includegraphics[width=8.5cm]{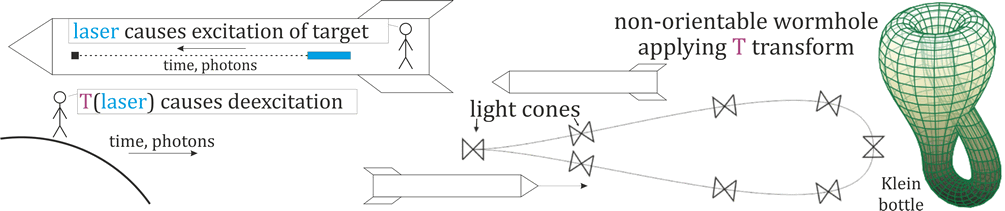}
        \caption{In theory, Einstein's general relativity allows for \href{https://en.wikipedia.org/wiki/Klein_bottle}{Klein-bottle}-like \href{https://en.wikipedia.org/wiki/Non-orientable_wormhole}{non-orientable wormholes}, travelling through which could apply T symmetry. Imagining a rocket travelled through it, from perspective of external observer, laser inside this rocket would \textbf{cause deexcitation of target} (stimulated emission, pull photons). Fig. \ref{cyclic} contains hypothetical cosmological alternatives for time reversal: before Big Crunch, Big Bounce. Section \ref{slaser} proposes more achievable realizations, applications of such hypothetical possibility. }       \label{Klein}
\end{figure}

While it is natural to push objects with photons, turns out there are also lots of successful approaches for \textbf{optical pulling} (e.g. \cite{pull}) starting with \href{https://en.wikipedia.org/wiki/Optical_tweezers}{optical tweezers}~\cite{tweezers} awarded the 2018 Nobel prize. Generally EM radiation pressure is a vector $\vec{p}=\langle \vec{E}\times \vec{H} \rangle/c$, allowing for \textbf{negative radiation pressure} to pull e.g. solitons~(\cite{neg1,neg2}). While there might also exist other approaches, here we propose a few laser settings, which assuming CPT symmetry should allow to create analogous \textbf{negative photon pressure} in narrow spectrum, and discuss their potential applications.

Lasers are believed to be governed by two basic equations for population of ground $N_1$ and excited $N_2$ state, for 
$$\textrm{stimulated emission:}\ \frac{\partial N_2}{\partial t}=-\frac{\partial N_1}{\partial t}=-B_{21}\,\rho(\nu) N_2$$
\be\textrm{and absorption:}\ \frac{\partial N_2}{\partial t}=-\frac{\partial N_1}{\partial t}=B_{12}\,\rho(\nu) N_1\ee
governed by $B_{12}=B_{21}$ \href{https://en.wikipedia.org/wiki/Einstein_coefficients}{Einstein coefficients}~\cite{einstein}. Both these equations act on e.g. pumped laser crystal, often with population inversion $N_2>N_1$. Additionally, produced photons are often absorbed by some external target, stimulating its excitation. Looking at it from perspective after CPT symmetry, stimulation equation should also apply to some external target, what means stimulated emission equation should act on it in standard perspective (no CPT). To exploit it, these should be different targets, also initially excited ($N_2>0$ e.g. lamp) - what requires some asymmetry, available for a few settings discussed further as in Fig. \ref{lasar}, e.g. Free Electron Laser, synchrotron, ring laser.
\subsection{Proposed realizations as lasAr (emission $\leftrightarrow$ Absorption)}
CPT theorem, originally proven by Julian Schwinger~\cite{CPT}, says that CPT symmetry (of charge conjugation + parity transformation + time reversal) holds for all physical phenomena. It allows to transform Feynman diagrams into their still valid CPT analogues: replacing particles with antiparticles, also reversing time. While it has lots of microscopic confirmations~\cite{CPTdata}, the big question is if it is still valid for macroscopic settings - if their CPT analogues should also work as supposed? In other words, if preparing CPT analogue of a scenario (e.g. initial conditions), should it lead to CPT symmetry of behavior, evolution? In theory, decomposing a setting into Feynman diagrams, and CPT transforming each of them, they should build CPT analogue of the original setting - which should work as in standard physics. Including causality: preparing CPT analogue of a scenario with clear causality, its time direction should be reversed.

\begin{figure}[t!]
    \centering
        \includegraphics[width=8.5cm]{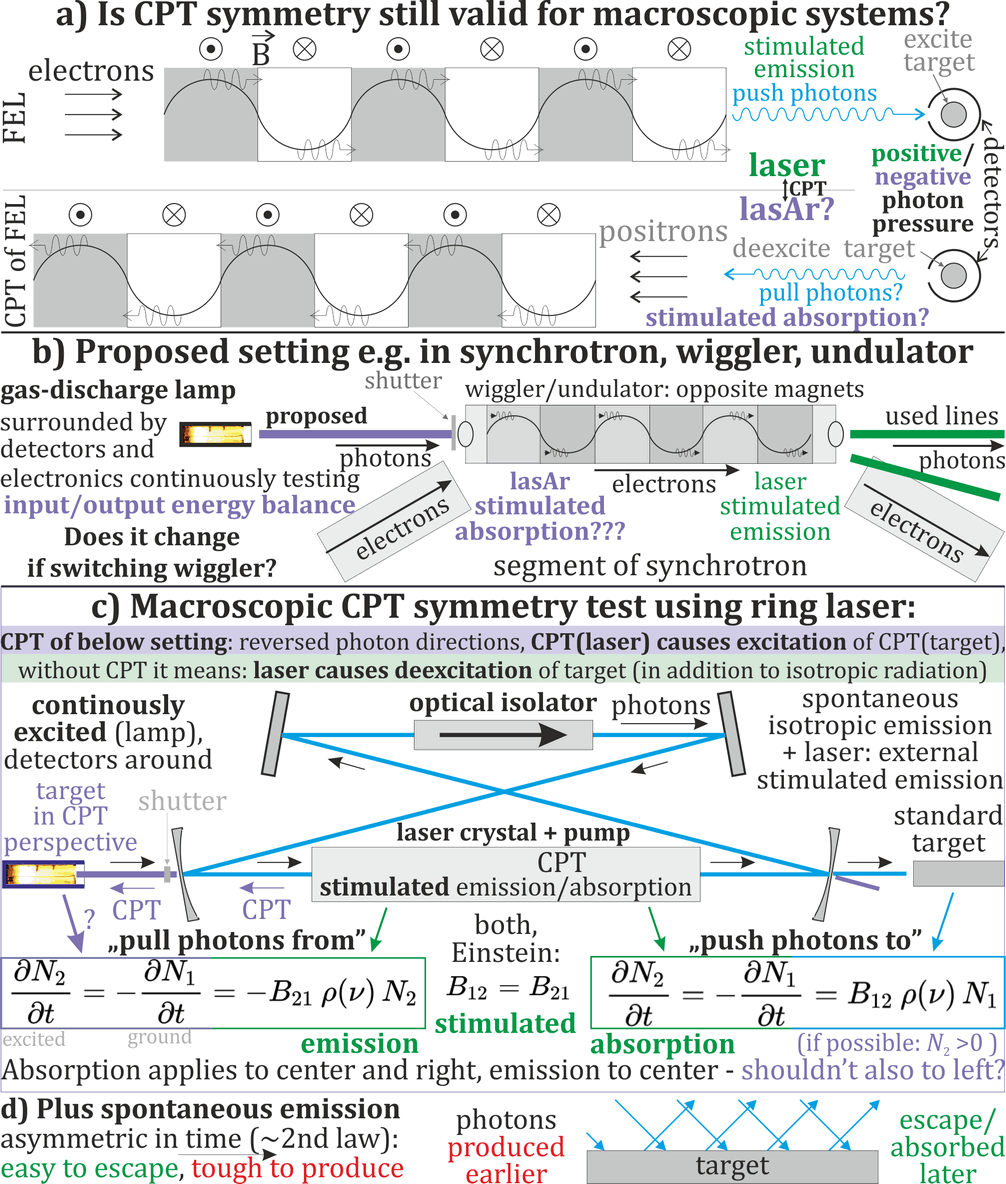}
        \caption{Potential consequences of hypothesis that CPT theorem is still valid in macroscopic physics. 
        a) In \href{https://en.wikipedia.org/wiki/Free-electron_laser}{free electron laser} (FEL) electrons travelling in properly shaped magnetic field emit photons as synchrotron radiation, \textbf{causing later excitation of target}. Performing its CPT transform, positrons travel in the opposite direction, \textbf{causing earlier deexictation of target} - getting \textbf{lasAr} replacing stimulated emission with stimulated absorption.
        b) Proposed experimental test using e.g. a \href{https://en.wikipedia.org/wiki/Gas-discharge_lamp}{gas-discharge lamp}, which excites atoms when powered - normally deexciting with spherically symmetric angle distribution (isotropic radiation). Placing it behind a working wiggler/undulator/FEL e.g. in synchrotron, the above CPT transform suggests it should additionally increase probability of deexcitation in this direction (in addition to isotropic) - what could be observed by surrounding the lamp with detectors (with exception to lasAr direction), and monitoring input/output energy balance of the lamp.
        c) more accessible setting: \href{https://en.wikipedia.org/wiki/Ring_laser}{ring laser} - using \href{https://en.wikipedia.org/wiki/Optical_isolator}{optical isolator} to enforce nearly unidirectional photon trajectories. From perspective after applied CPT symmetry, photon trajectories would be reversed - exciting the target, what from standard perspective (no CPT) means causing deexictation of target. From perspective of written standard \href{https://en.wikipedia.org/wiki/Stimulated_emission}{stimulated emission} equations, both act on the central target (pumped laser crystal). Additionally, the laser "pushes" photons to target on the right, accordingly to the absorption equation. In perspective after CPT symmetry, this equation would apply to target on the left, what without CPT means that the emission equation applies to it - that laser "pulls photons from" this target if possible: it was initially excited $(N_2>0)$. e.g. as gas-discharge lamp. Its excited atoms would spontaneously deexcite, producing isotropic radiation.
        d) In contrast to stimulated emission/absorption, spontaneous processes are time asymmetric due to solution we live in: it is easy for photons to escape to the future, it is difficult to produce them in the past.
}       \label{lasar}
\end{figure}

Building CPT analogue of a scenario seems technically extremely challenging, however, there might be found situations where it is technically reachable. Free Electron Laser (FEL) seems to be such an example (also just synchrotron radiation). It is just electrons travelling in properly shaped magnetic field, through synchrotron radiation producing photons, which finally cause excitation of the target - as in Figure \ref{lasar}. From perspective after CPT transform: initially excited target produces photons, finally absorbed by positrons travelling in the opposite direction. 

From causality perspective, \textbf{FEL through stimulated emission causes later excitation of target}, while in its CPT transformed analogue: \textbf{stimulated absorption causes earlier deexcitation of target}. In other words, at least naively, laser should become \textbf{lasAr} - replacing stimulated emission with stimulated absorption: $\textrm{laser} \overset{\textrm{CPT}}{\leftrightarrow} \textrm{lasAr}$. If the target is gas-discharge lamp (of overlapping spectrum), it would naturally deexcite in isotropic way - the question to test is if such lasAr could additionally slightly increase probability of deexcitation of these atoms in its direction and spectrum? As in Fig. \ref{Klein}, stimulation of target deexcitation seems also allowed by Einstein's general relativity.

The use of positrons does not seem to matter, suggesting to just use electrons instead - going back to FEL. It also brings question why such effects are not already observed there? The answer seems simple: because it would require target previously excited in the corresponding spectrum, what seems usually not satisfied, and generally not searched for - but should be relatively easy to test in a dedicated experiment e.g. in synchrotron or FEL. 

While such CPT analogue of laser was proposed by the author in 2009 (e.g. \url{https://groups.google.com/forum/#!topic/sci.physics.foundations/xhUfe8akaS0}), it seems still remain untested experimentally, hopefully to be repaired in a near future e.g. in \textbf{synchrotron} like Solaris in Cracow. As b) in Figure \ref{lasar}, it seems to just require placing a gas-discharge lamp behind a segment (necessary transparent window, alternatively behind the bending magnet/main synchrotron photon source), surrounded by detectors (with hole toward lasAr) combined with electronics monitoring input/output power balance of lamp-detectors. Activating the wiggler/undulator/FEL targeting this lamp, naively should increase probability of deexcitation in this direction (in addition to isotropic radiation), what should be seen as lowered energy in the monitored energy balance.

Such test has turned out technically challenging (far ultraviolet, no FEL), however, a colleague has pointed existence of \textbf{ring lasers}, suggesting more accessible alternative as in Figure \ref{lasar}. They use closed photon trajectories, a single direction can be made dominating inserting optical isolator. From perspective after CPT transform, this direction would be reversed (also switching written two equations) - causing excitation of the shown target "behind the laser", what in the original perspective (no CPT) would mean \textbf{causing deexcitation of this target} - seen as darkening by detectors around (e.g. a camera with filter). From equations perspective ($N_2$ is number of excited atoms), $\partial N_2/\partial t = -B_{21} \rho(\nu) N_2$ should act not only on the central target, but (symmetrically  to the second equation) also on the target on the left, if only it is initially excited: $N_2>0$, the more the stronger the effect. Performing such test seems much more accessible, hopefully to be conducted in a near future. However, in contrast to FEL/synchrotron, the asymmetry is imperfect here - there is a percentage of photons travelling in the opposite direction, which would excite the target - against the hypothetical stimulated deexcitation, making it more difficult to detect. In this case, e.g. fast rotation of target might allow to distinguish the two effects. 

\begin{figure}[t!]
    \centering
        \includegraphics[width=8.5cm]{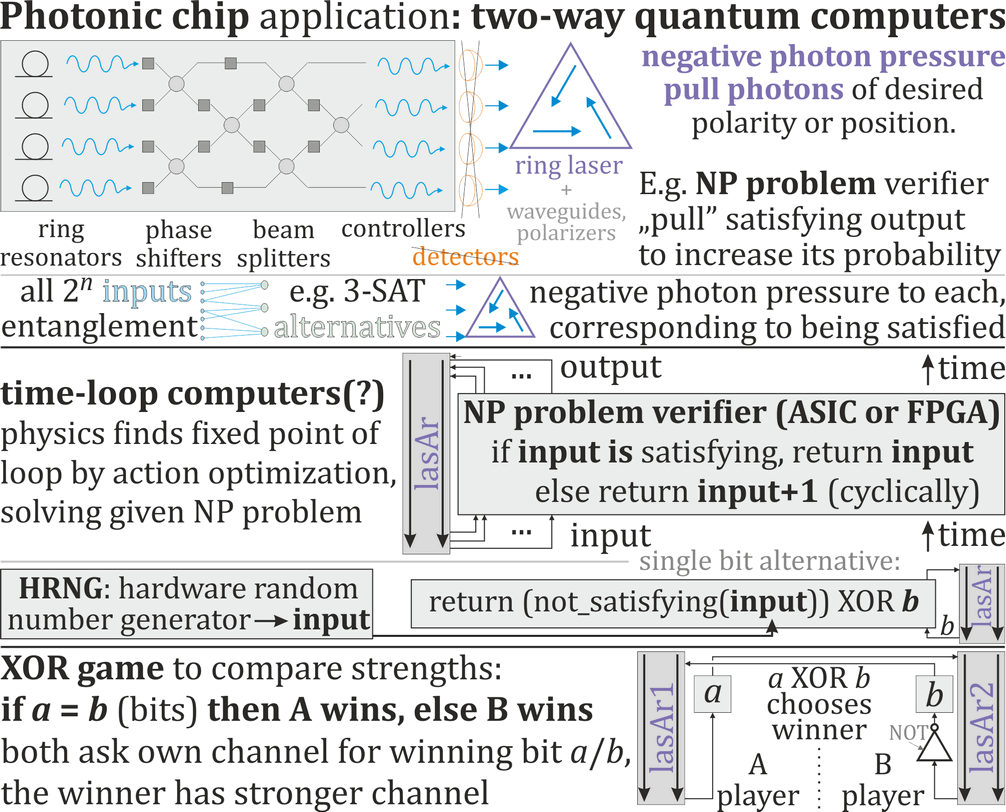}
        \caption{Some potential computational applications of proposed lasAr as CPT analogue of laser (e.g. free electron laser, synchrotron, ring laser), hopefully creating negative radiation pressure (\cite{neg1,neg2}) for photons in narrow spectrum. \textbf{Top}: while Shor-like \href{https://en.wikipedia.org/wiki/One-way_quantum_computer}{quantum computers are one-way}: mounted only in the past by state preparation, e.g. Ising model based computer like in Fig.\ref{fqc} is two-way: mounted in both directions. LasAr e.g. ring laser might allow to mount in the future, replacing measurement by detectors, especially for photonic computers~\cite{photonic}. This way hopefully increase probability of producing photons of chosen e.g. polarity. For example implementing verifier of NP problem, and pulling photons corresponding to its satisfactory results (e.g. 3-SAT split into alternatives), hopefully to restrict initial entanglement of all inputs to the satisfying ones.        
        \textbf{Center}: potential applications for \textbf{time-loop computers}: making that \href{https://en.wikipedia.org/wiki/Stationary-action_principle}{action optimization} finds fixed point of a (time) loop, designed to solve given instance of \href{https://en.wikipedia.org/wiki/NP_(complexity)}{NP problem} this way. To send multiple bits with such lasAr, one could use spectral, temporal or spatial division, e.g. lattice of shutters (or mirrors like in DLP projectors). There is also shown alternative sending single bit - creating contradictory NOT time loop, unless HRNG generates satisfying input - hopefully preferred for action optimization. 
        \textbf{Bottom}: Constructed time loops might be self-contradictory, requiring imperfections: possibility to lie, breaking such causal loop, e.g. inside the used electronics or such channel back in time. The proposed XOR game allows to compare strength of such hypothetical channels back in time: player A chooses bit $a$, player B chooses bit $b$, and they have opposite goals. They use own channels to send the final values, optimizing own goals - the one with stronger channel should win. 
        }       \label{comp}
\end{figure}

\subsection{Potential applications of negative photon pressure}
If such test will turn out successful, or there are found other realisations of negative photon pressure, applications could start e.g. with increasing rate of some chemical (e.g. photolitography) or nuclear transitions producing  characteristic photons - "pulling photons" of this energy by placing such target behind e.g. ring or free electron laser tuned to this energy. Much more ambitious future application might be stimulated proton decay (if it is possible, energy density $>100\times$ than fusion, from any matter) by some optimized sequence of photon pushing and pulling. 

The most crucial direction seems potential information/computation applications, summarized in Fig. \ref{comp}. The basic one is hypothetical, hopefully allowing to solve especially NP complete problems, \textbf{two-way quantum computer} enhancement - mounted in both directions as e.g. Ising computer in Fig. \ref{fqc} by some CPT analogue of state preparation. For example photonic quantum computer with detectors replaced e.g. with ring laser pulling photons of chosen polarity - in which there are encoded e.g. results of alternatives for 3-SAT problem, adding preference to choose a satisfying input by their initial entanglement from some Hadamard gates. If it is possible, the effect might be small change of measurement probability distributions, which could be strengthen e.g. by some error correction.

Adding shutter to lasAr, CPT arguments suggest the stimulated emission in target naively should happen earlier than opening shutter by optical path length: distance divided by speed of light ($\approx 3.3$ nanoseconds per meter). If it will turn out true, and overcoming delays of the setting/electronics, it might allow for information channel back in time between shutter and electronics monitoring energy balance of the lamp. Adding some delay line, e.g. with mirrors, optical fibers, techniques for slowing down light, might allow to overcome the delays - to literally send one bit of information slightly earlier. To extend it to multiple bits, one could e.g. use spatial division (like lattice of shutters or mirrors as in DLP projector), frequency division, maybe also temporal division (more complicated here). 

Achieving nanosecond-scale time difference this way would be sufficient to put a simple \href{https://en.wikipedia.org/wiki/Application-specific_integrated_circuit}{ASIC chip} in time loop (e.g. 3-SAT verifier): send its output back, and use as its earlier input. Building such ASIC chip for a given instance of NP-complete problem (e.g. 3-SAT), as in Fig \ref{comp}: output the input if it is correct, or input+1 (cyclically) otherwise, would transform this problem into search of fixed point of this loop. Closing this loop with cables, if there is clock it would check one input per cycle until finding a correct one. It is interesting open question what would happen without clock - some complex electron hydrodynamics which should stabilize by finding the fixed point of the loop (solving our problem). Closing this loop in time e.g. by lasAr, action optimization of Lagrangian mechanics governing our world should find this fixed point - solving the given computational problem ... unless there is a simpler way to break such loop, e.g. lie in such channel or electronics.

Self-contradictory loops are also possible, e.g. with just NOT gate - leading to observed oscillations if being closed in space. Being able to close NOT into a loop in time, action optimizing physics would have to make one of them (channel or gate) to lie. Therefore, such channels cannot be perfect, always need to have a way to lie - breaking the weakest link (for action optimization) of given self-contradictory causal loop. For this purpose action optimization could e.g. exploit imperfections of the used electronics, however, such hypothetical channel seems the most likely weakest link. To test, compare strength of such channels as capability to avoid lying in difficult cases, Fig. \ref{comp} also suggests XOR game: putting them in situation in which exactly one is right, the stronger channel should win (alternatively e.g. NOT gate might lie). For such evaluation there could be alternatively used some standardized chips with easy to break causal links. To increase channel strength there could be used higher intensity lasers, more saturated targets (high $N_2$), combined multiple channels, error correction techniques, etc.

While the mentioned time-loop computer approach would require sending multiple bits, alternative approach sending single bit $b$ could be: use a good hardware random number generator (HRNG, e.g. quantum measurement) to choose input sent to the verifier, which also sends back in time to itself: bit $b$ if input is satisfying, NOT($b$) otherwise. This way to avoid contradictory NOT time loop, action optimization should make this generator already choose a satisfying input. Generally making choices based on good HRNG could give action optimization freedom to optimize its generated values based on their later consequences, also in potential macroscopic applications.

Assuming CPT remains valid for macroscopic physics, still construction of such time-loop computer would be technically extremely challenging (strong channels, error resistant electronics), but it might be reachable - should be at least taken into considerations. Especially that it might allow to break currently used cryptography (if reaching sufficient delay and strength): with verifier checking if a given key leads to decoded file not being just a noise. Protection against such attacks might be done by adding computationally costly initialization: necessary before application of a new cryptographic key - to require much longer delays and stronger channels. However, it might finally lead to post-cryptography world: with safe only basic techniques like one-time pad. Further improvements of such channels could allow to use verifier e.g. testing molecules for desired properties for drug design, testing possible algorithms/methodologies/procedures for desired outcomes, etc. - to find satisfying choices with one of such time-loop approaches for problems testable in practical time, as physical world analogies of NP problems.

Finally, being able to send information back for macroscopic time differences (e.g. using systems of satellites for delay lines), could allow to prevent currently unpredictable unwanted events: there would be attempt to send back such missing information - creating inconstancy, hence action optimization should modify the weakest links of such reason-result chain (e.g. in quantum measurement level) to get self-consistent time loop (\href{https://en.wikipedia.org/wiki/Novikov_self-consistency_principle}{Novikov self-consistency principle}) - for example with satisfying outcome: not requiring to use such channel. In other words, just having access to such channel, we would enforce physics to make its best to prevent our bad choices, optimize randomness for outcomes preferred by us - like mentioned creation of contradictory NOT time loop for non-satisfying outcomes (easier to optimize if leaving freedom by making choices based on good HRNG). While there would be many dangers on the way, if well balancing strengths of such channels worldwide (between players of different/opposite goals), it might lead to a much more harmonic world based on trust and common goals, potentially without crime and various types of gambling (economy based on objective values), with choices made optimizing their actual future consequences - maybe also avoiding wars, suboptimal politicians, dangers of new technologies, etc.

\section{Conclusions and further perspectives}
While our natural intuition suggests us "evolving 3D" picture of the world: which e.g. allows to conclude Bell inequalities for resulting correlations, quantum mechanics has Born rules instead: squares relating amplitudes and probabilities - leading to violation of such inequalities. Living in 4D spacetime, what is nonintuitive but required by modern understanding of physics like special relativity, Lagrangian mechanics or QFT, requires to consider paths/scenarios (Feynman diagrams) as the basic objects - what leads to quantum-like behavior, starting with Anderson localization, Born rules and Bell violation. Specifically, its consequence is the present moment being in equilibrium between past and future, tension from both these directions is described by the asymptotic behavior for example of $M^t=(-\mathcal{H})^t$ propagator from infinity as in (\ref{pathen}) (left/right in Ising model), which is quantum amplitude of the ground state. Finally to randomly get some value of a measurement, we intuitively need to draw it from both time directions, getting probability being the square of amplitude. Consequences of living in 4D spacetime are seen in many quantum experiments like Wheeler's, delayed choice quantum erasure, as discussed also in Shor's algorithm, maybe to be extended to even stronger time-loop computers.

Another conclusion from MERW is that the stochastic and quantum realms of physics, which have historically split their ways for example due to disagreement in predictions for semiconductor due to Anderson localization, can reunite if repairing the subtle approximation in entropy maximization, using ensembles of full paths. Especially interesting and important are situations in intersection of both worlds, like good understanding of electron flow in microscopic systems, what is crucial in modern electronics reaching level of single atoms and molecules, where standard Ohm's law is not satisfied: a fixed potential difference for identical local situations can lead to different currents, behaving in a nonlocal way: depending on the entire system like in Fig. \ref{cond}. MERW-based modelling can be used as a practical approximation of extremely costly complete quantum calculations e.g. of p-n junction - discussed in \cite{cond1}. \\

\begin{figure}[b!]
    \centering
        \includegraphics{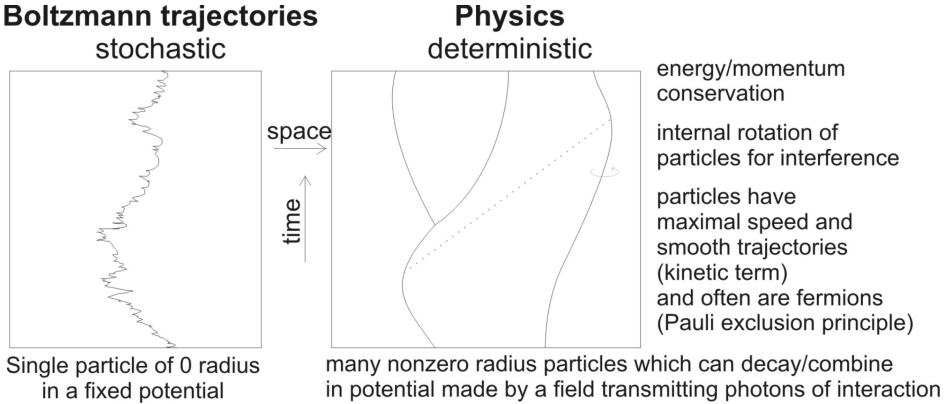}
        \caption{Boltzmann distribution among trajectories is a simple effective model recreating some basic properties of physics, quantum mechanics. One expansion toward fundamental physics is adding possibility to combine or decay particles, requiring to extend the ensemble of objects to consider from paths to more complex scenarios like Feynman diagrams. This way we would go to universal picture of perturbative quantum field theories. Further steps toward fundamental physics will eventually require asking for field configurations behind such single diagrams, including gluing of EM fields in the center of charge, avoiding energy infinity obtained by assuming that charge is a perfect point. Such localized field configurations for particles are technically called solitons, for which quantization of charge has well known mathematical analogue: topological charge, discussed in Fig. \ref{charge}. }
       \label{furth}
\end{figure}

\begin{figure}[b!]
    \centering
        \includegraphics{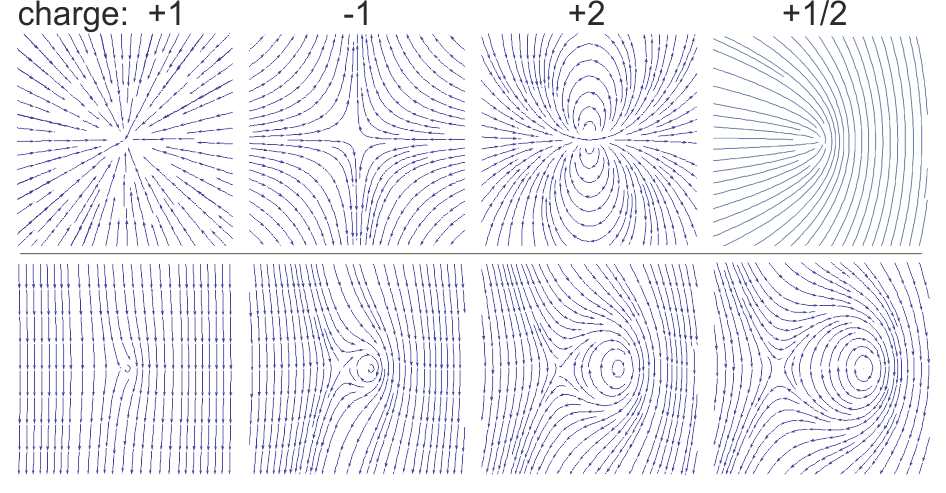}
        \caption{Top: examples of 2D topological charges of vector field - looking at a loop around such singularity, charge is the (integer) number of complete field rotations. If "removing arrows of vectors" we can also get charges being multiplicity of 1/2 this way (like spin). Well known example in nature of such field configuration stabilized by topology (topological soliton) is fluxon (Abrikosov vortex) in superconductor, carrying quant of magnetic field. Going to 3D analogs, e.g. in liquid crystals there are observed also Coulomb-like interactions~\cite{coulomb} for such topological charges. Mathematically, defining EM field as curvature of vector field, Gauss-Bonnet theorem acts exactly as the Gauss law: integrating curvature  over a closed surface, we get the total topological charge inside this surface - leading to electrodynamics with Maxwell's equations governing dynamics of such quantized charges~\cite{faber}. }
       \label{charge}
\end{figure}

As discussed, Boltzmann distribution among paths is a simple model suggested by many perspectives, like being successfully used in Ising model, the principle of maximum entropy of statistical physics, living in 4D spacetime, or agreement with expected quantum predictions and confirming experiments - making it a promising direction for understanding of physics governing our world. However, it obviously also contains essential simplifications, some of which are visualized in Fig. \ref{furth}. One of them is varying number of particles in real physical scenarios, what is repaired in perturbative quantum field theories by considering ensemble among more sophisticated scenarios: Feynman diagrams. A real scenario represented by such simplified diagram contains additionally configuration of fields of interactions, for example electromagnetic, suggesting field theories for more fundamental description. Particles having a charge maintain nearly singular configuration of electric field - robust configurations of fields are technically called solitons, topology brings a natural mathematical tool to explain their charge quantization like in Fig. \ref{charge}.

One of essential properties ignored by Boltzmann distribution among paths is the requirement for interference: some particle's internal periodic process, called de Broglie's clock ($E=mc^2=\hbar\omega$) or zitterbewegung, which has been directly observed for electron for example as increased absorption when synchronizing period of such clock with lattice constant of silicon crystal~(\cite{hest, elclock}). It causes coupled "pilot waves" of the surrounding field, confirmed as de Broglie-Bohm interpretation of QM for example by experiment measuring average trajectories of interfering photons in double-slit experiment~\cite{average}. While principle of complementarity forbids measuring both corpuscular and wave natures simultaneously, it does not exclude that particle has objectively both natures at a time, especially that no conditions for choosing one of them are specified (e.g. in which moment meeting electron and proton become hydrogen?), or mechanisms for such change of nature, and there are experiments successfully exploiting both natures at a time like Afshar's~\cite{afshar}.

There are also lots of experimental hydrodynamical analogs of QM, especially started by Yves Couder, which show that such classical objects coupled with waves they create (droplet on a vibrating liquid surface) allow to recreate many quantum phenomena like: interference~\cite{cd1} in double-slit experiment (particle goes one trajectory, interacting with waves it created - going through all trajectories), tunneling~\cite{cd2} (depending on practically random hidden parameters - highly complex state of the field), orbit quantization~\cite{cd3} including double quantization~\cite{cd5}: of both radius and angular momentum like in Bohr-Sommerfeld (particle has to "find a resonance" with the field - its internal phase has to return to initial state during full orbit), Zeeman splitting analogue~\cite{cd4} (using Coriolis as Zeeman force), like in MERW: recreating quantum eigenstates with statistics of trajectory~\cite{cd6}, also Bell violation~\cite{belldrop}.

\begin{figure}[t!]
    \centering
        \includegraphics{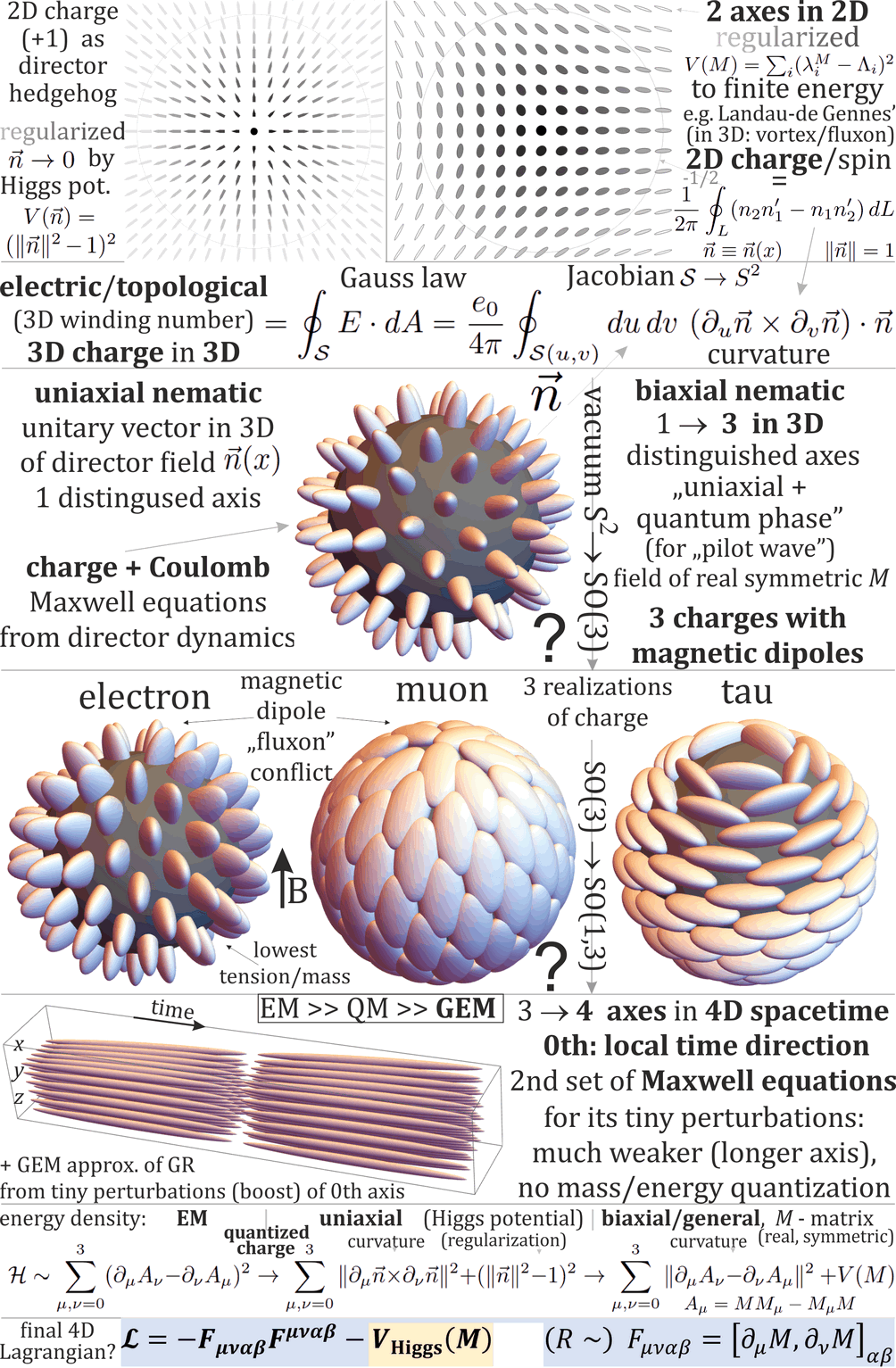}
        \caption{Topological soliton particle model approach~\cite{biaxial} (gathered materials: \url{https://github.com/JarekDuda/liquid-crystals-particle-models}) exploring resemblance of topological defects in liquid crystals, e.g. experimentally allowing to obtain Coulomb-like interaction~\cite{coulomb}. Higgs-like potential enforces e.g. vector field to of unitary vectors far from particles, also allowing to regularize singularity (charge) to finite energy. \textbf{Interpreting field curvature as EM field, Gauss law returns topological charge} - getting Maxwell equations with built-in charge quantization (Faber's apprach~\cite{faber}). In biaxial nematic, governed by modification of \textbf{Landau-de Gennes model}, we have 3 distinguishable axes, leading to 3 types of hedgehog: the same charge, but different mass - resembling 3 leptons, also requiring magnetic dipole moment due to the hairy ball theorem. Extending further to field of 4 distinguishable axes in 4D ($SO(3) \to SO(1,3)$ vacuum), represented by real symmetric tensor field ($M$), leads to additional second set of Maxwell equations - as in gravitoelectromagnetism approximation of the general relativity theory.}
       \label{biaxial}
\end{figure}

\begin{figure}[t!]
    \centering
        \includegraphics{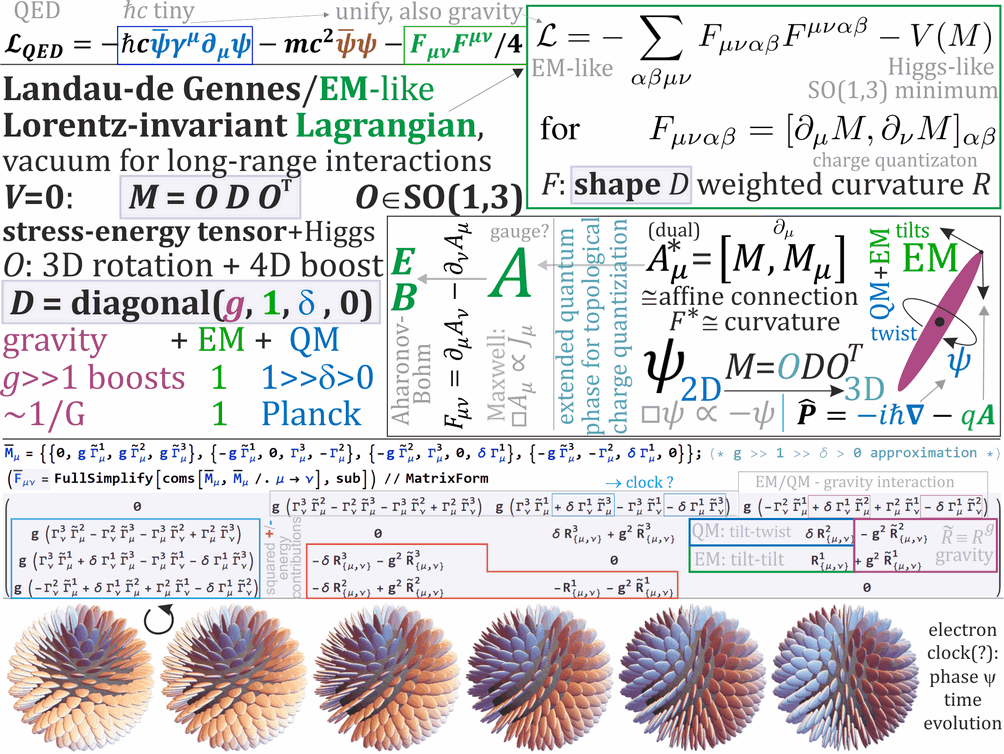}
        \caption{Top right: Lagrangian candidate for model in Fig. \ref{biaxial}. It is based on Landau-de Gennes model, but with Lagrangian using $F_{\mu\nu}F^{\mu\nu}$ characteristic EM term - this time with $F$ interpreted as curvature of a deeper real symmetric field $M$ for topological charge quantization, and using its full 4 index form - for unification with the remaining wave-like dynamics: of quantum phase ($\sim$Klein-Gordon, first term in QED Lagrangian) and gravity in GEM approximation. Unification of these 3 different long-range interactions, differing by many orders of magnitude, is made through the shape $D$ of rotated object ("molecule" in liquid crystal analogy) - energetically preferred eigenvalues of $M$ field  due to Higgs-like potential. SO(3) vacuum unifies quantum phase (low energy twists), and $S^2$ for electromagnetism - Gauss law counts its topological charge for quantization, then SO(1,3) adds boosts with dynamics corresponding to \href{https://en.wikipedia.org/wiki/Gravitoelectromagnetism}{gravitoelectromagnetic} approximation of the general relativity.  }
       \label{lagr}
\end{figure}

\begin{figure}[t!]
    \centering
        \includegraphics{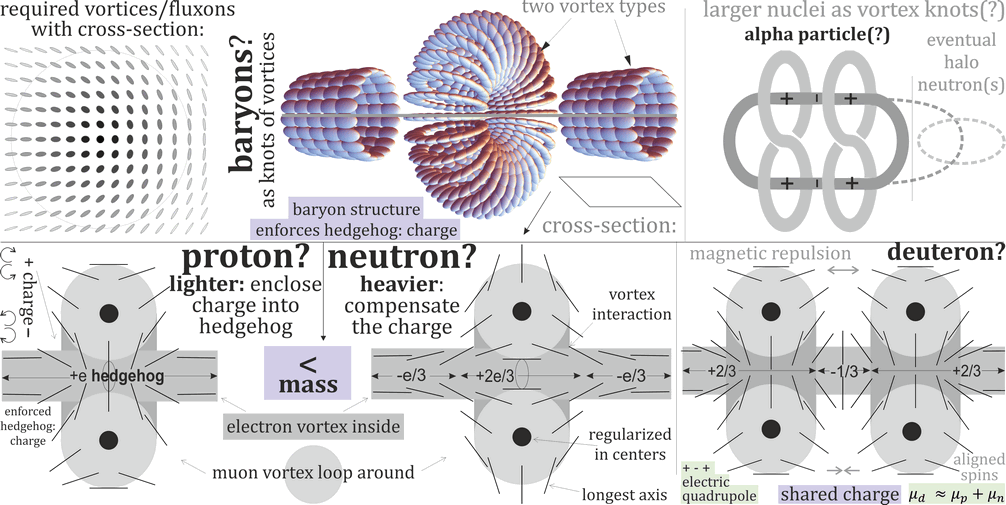}
        \caption{Further topological defects in model from Fig. \ref{biaxial} resemble e.g. baryons, with intuitive explanation \textbf{why proton is lighter than neutron}. This model requires fluxon-like topological vortices, which shown simplest knot has interaction between two vortices - we can see the external one enforces hedgehog-like configuration in the central one. Full hedgehog is topological charge corresponding to electric charge - proton can just enclose to. In contrast, neutron has to compensate this charge, what requires additional mass/energy. Such mechanism of \textbf{baryons requiring some charge} 
        can also explain e.g. deuteron binding - sharing single charge to reduce energy, also explaining known electric quadrupole moment, and that spins are aligned: $\mu_d\approx \mu_p +\mu_n$. The vortices allow here to bind larger nuclei against Coulomb repulsion, including neutrons in large distance in \href{https://en.wikipedia.org/wiki/Halo_nucleus}{halo nuclei}. As there is no Gauss law for baryon number, this model allows for its violation, as e.g. in \href{https://en.wikipedia.org/wiki/Baryogenesis}{baryogenesis} or \href{https://en.wikipedia.org/wiki/Hawking_radiation}{Hawking radiation}. Being able "to swing" such configuration (e.g. with lasers) out of such very deep local minimum (to unknot in this model), what could be optimized with such model, might allow for hypothetical stimulated proton decay. }
       \label{baryon}
\end{figure}

The universality of quantum formalism is also recreated in other hydrodynamical analogues, e.g. Casimir effect: two plates in vibrating water tank also were experimentally shown to attract~\cite{casimir} as wave energy between them is lower due to restriction by the plates. There is also hydrodynamical analogue of Aharonov-Bohm effect suggested by Berry~\cite{berry}: using vortex, vorticity and Casimir force analogues of solenoid, magnetic field and Lorentz force.

To summarize, while there is unsuccessful belief that we need to find a boundary between classical and quantum world, this boundary blurs e.g. with hydrodynamical analogues or MERW - it might turn out nonexistent: they can be just different perspectives/descriptions of the same system. For example \href{https://en.wikipedia.org/wiki/Normal_mode#Coupled_oscillators}{coupled oscillators} can be described by evolution of their positions ("classical"), or in the base of their normal modes - where this evolution becomes literally unitary ("quantum"). Lattices of such oscillators are used to model crystals: classically, or equivalently in Fourier basis: using phonons as normal modes - which are treated as (quasi)particles in Feynman diagrams. In continuous limit of such lattice we get field theories - which can be modelled with hydrodynamical analogues. Solitons are localized particle-like configurations of fields, effectively described by QFT. Using topological solitons we get charge/spin quantization, pair creation/annihilation, and electromagnetism-like interaction for them. Finally Couder's quantization suggests how to understand atoms: Schr\"{o}dinger equations describes coupled wave, which to minimize energy needs to become a standing wave - this resonance between particle's clock and the field gives quantization conditions. Fig. \ref{charge}, \ref{biaxial}, \ref{lagr}, \ref{baryon} present some basics of such particle approach, discussed e.g. in \cite{faber,biaxial}.

\bibliographystyle{IEEEtran}
\bibliography{cites}
\end{document}